\documentclass[11pt,
               abstract=on
]{scrartcl}

\usepackage[a4paper, left=30mm, right=30mm, top=20mm, bottom=30mm]{geometry}

\usepackage[ruled]{algorithm2e}

\usepackage{booktabs}
\usepackage{multirow, array}
\usepackage{graphicx}
\usepackage{caption}
\usepackage{subcaption}

\usepackage{xcolor}
\definecolor{dodgerblue}{RGB}{30,144,255}
\definecolor{steelblue}{RGB}{70,130,180}
\definecolor{darkred}{RGB}{178,34,34}
\definecolor{forestgreen}{RGB}{34,139,34}

\usepackage[pagebackref]{hyperref}
\hypersetup{
    colorlinks=true,
    citecolor=blue, 
    citecolor=steelblue,
    linkcolor=forestgreen
    }

\usepackage[auth-sc, affil-it]{authblk} 

\usepackage{amsmath, amssymb, amsthm, amsfonts} 
\usepackage{bm} 
\usepackage{bbm}
\usepackage{csquotes}
\usepackage{todonotes}
\usepackage[round]{natbib}
\newcommand{\bx}{\mathbf{x}}
\newcommand{\by}{\mathbf{y}}
\newcommand{\bs}{\mathbf{s}}

\newcommand{\bth}{\bm{\theta}}
\newcommand{\bTh}{\bm{\Theta}}
\newcommand{\bpsi}{\bm{\psi}}

\renewcommand{\d}{\mathrm{d}}
\newcommand{\mesh}{\mathrm{mesh}(0,T)}
\newcommand{\Ds}{\mcal{D}_{\textsc{s}}}
\newcommand{\Dm}{\mcal{D}_{\textsc{m}}}

\newtheorem{Theorem}{Theorem}
\newtheorem{Proposition}{Proposition}
\newtheorem{Definition}{Definition}
\newtheorem{Remark}{Remark}
\newtheorem{Lemma}{Lemma}
\newtheorem{Example}{Example}
\newtheorem{Corollary}{Corollary}

\usepackage[acronym,smallcaps,nowarn,section,
nonumberlist,shortcuts]{glossaries}
\glsdisablehyper
\newacronym{abc}{abc}{approximate Bayesian computation}
\newacronym{lfi}{lfi}{likelihood-free inference}
\newacronym{gbi}{gbi}{generalised Bayesian inference}
\newacronym{saabc}{sa-abc}{semi-automatic \textsc{abc}}
\newacronym{mmd}{mmd}{maximum mean discrepancy}
\newacronym{mcmc}{mcmc}{Markov chain Monte Carlo}
\newacronym{skrr}{sr-abc}{Signature regression \textsc{abc}}
\newacronym{swd}{swd}{sliced Wasserstein distance}
\newacronym{k2abc}{k2-abc}{double kernel \textsc{abc}}
\newacronym{rej}{rej-abc}{rejection \textsc{abc}}
\newacronym{gbm}{gbm}{geometric Brownian motion}
\newacronym{mh}{mh}{Metropolis-Hastings}
\newacronym{ma2}{ma(2)}{moving average model of order 2}
\newacronym{wass}{wass}{Wasserstein distance}
\newacronym{sabc}{s-abc}{Signature \textsc{abc}}
\newacronym{iid}{iid}{independent, identically distributed}
\newacronym{pmcmc}{pmcmc}{particle Markov chain Monte Carlo}
\newacronym{rkhs}{rkhs}{reproducing kernel Hilbert space}
\newacronym{wl}{wl}{Weisfeiler-Lehman}
\newglossaryentry{cde}
{
  name={\textsc{cde}},
  description={controlled differential equation},
  first={\glsentrydesc{cde} (\glsentrytext{cde})},
  plural={\textsc{cde}s},
  descriptionplural={controlled differential equations},
  firstplural={\glsentrydescplural{cde} (\glsentryplural{cde})}
} 
\newglossaryentry{sde}
{
  name={\textsc{sde}},
  description={stochastic differential equation},
  first={\glsentrydesc{sde} (\glsentrytext{sde})},
  plural={\textsc{sde}s},
  descriptionplural={stochastic differential equations},
  firstplural={\glsentrydescplural{sde} (\glsentryplural{sde})}
} 
\newcommand{\bd}[1]{\mathbf{#1}}
\newcommand{\bv}[1]{BV(\left[0,T\right], #1)}
\newcommand{\Pl}[1]{\mathcal{P}(\left[0,T\right], #1)}
\newcommand{\mcal}[1]{\mathcal{#1}}
\newcommand{\sig}[1]{\mathrm{Sig}(#1)}
\newcommand{\kap}[1]{\kappa(#1, \cdot)}
\newcommand{\onevar}[1]{\| #1 \|_{\textrm{1-var}}}
\newcommand{\norm}[2]{\left\lVert#1\right\rVert_{#2}}
\newcommand{\pH}{\prod_{m \geq 0} \mcal{H}^{\otimes m}}
\newcommand{\ip}[3]{\left\langle #1, #2 \right\rangle_{#3}}

\newcommand\blfootnote[1]{%
  \begingroup
  \renewcommand\thefootnote{}\footnote{#1}%
  \addtocounter{footnote}{-1}%
  \endgroup
}

\usepackage{listings}

\definecolor{codegreen}{rgb}{0,0.6,0}
\definecolor{codegray}{rgb}{0.5,0.5,0.5}
\definecolor{codepurple}{rgb}{0.58,0,0.82}
\definecolor{backcolour}{rgb}{0.95,0.95,0.92}

\lstdefinestyle{mystyle}{
  backgroundcolor=\color{backcolour}, commentstyle=\color{codegreen},
  keywordstyle=\color{magenta},
  numberstyle=\tiny\color{codegray},
  stringstyle=\color{codepurple},
  basicstyle=\ttfamily\footnotesize,
  breakatwhitespace=false,         
  breaklines=true,                 
  captionpos=b,                    
  keepspaces=true,                 
  numbers=left,                    
  numbersep=5pt,                  
  showspaces=false,                
  showstringspaces=false,
  showtabs=false,                  
  tabsize=2
}

\newcolumntype{C}{>{\displaystyle}c}

\subject{\footnotesize Preprint}

\title{Approximate Bayesian Computation\\ with Path Signatures}

\author[1,2]{\small Joel Dyer\textsuperscript{*}}
\author[3]{\small Patrick Cannon}
\author[4]{\small Sebastian M Schmon}

\affil[1]{{\normalfont Mathematical Institute, University of Oxford, UK}}
\affil[2]{{\normalfont Institute for New Economic Thinking, Oxford}}
\affil[3]{{\normalfont Techwerk}}
\affil[4]{{\normalfont Department of Mathematics, Durham University, UK}}

\date{\small \today}

\begin{document}

\maketitle

\begin{abstract}
{\small
\noindent Simulation models often lack tractable likelihood functions, making likelihood-free inference methods indispensable. Approximate Bayesian computation generates likelihood-free posterior samples by comparing simulated and observed data through some distance measure, but existing approaches are often poorly suited to time series simulators, for example due to an independent and identically distributed data assumption. In this paper, we propose to use path signatures in approximate Bayesian computation to handle the sequential nature of time series. We provide theoretical guarantees on the resultant posteriors and demonstrate competitive Bayesian parameter inference for simulators generating univariate, multivariate, irregularly spaced, and even non-Euclidean sequences.
}
\end{abstract}


\vspace{-0.5cm}
\section{Introduction}
Simulation models are an increasingly popular tool in a broad range of scientific disciplines including cosmology \citep{Alsing_2018}, economics \citep{Geanakoplos2012}, and the biological sciences \citep{Christensen2015}. 
A drawback of such models is that, while they are straightforward to sample from, their complexity typically does not allow for explicit evaluation of the associated likelihood function. 
Consequently, traditional approaches to statistical inference are infeasible and alternative \gls{lfi} methods are usually adopted.
\blfootnote{\textsuperscript{*}
Email: \texttt{dyer@maths.ox.ac.uk}}

Many such approaches have been proposed. One of the most widely used \gls{lfi} methods is \gls{abc} \citep{tavare1997inferring, pritchard1999population, beaumont2002approximate}, in which the Bayesian posterior distribution is approximated by sampling parameters $\bth$ from a prior distribution and synthetic datasets $\bx$ from a stochastic simulator -- with likelihood denoted $p(\bx \mid \bth)$ -- and comparing the output $\bx$ with real data $\by$. If the simulator output is  sufficiently `close' to the observation, then $\bth$ is retained as a sample from the approximate posterior distribution; otherwise, it is discarded.

However, measuring closeness between model outputs is known to be challenging. This is particularly the case for time series data, which can exhibit complex dependency structures and may be multivariate and sampled at irregular time intervals. 
A common approach is to attempt to distil important features of the data using summary statistics and compare these instead \citep[see e.g.][]{prangle2018handbook}.
In practice, informative summary statistics are difficult to craft, which presents a trade off---a poor choice can materially bias the algorithm away from the true posterior distribution, yet constructing a sufficiently powerful choice can require substantial domain expertise, problem insight, and costly experimentation \citep[see e.g.][for a recent comparison of methods with and without summaries]{drovandi2021comparison}.

In other approaches the engineering of summary statistics is bypassed altogether in favour of distances on the full dataset \citep[e.g.][]{Park2016, jiang2018approximate, Bernton2019, nguyen2020approximate}. 
However, in many such cases the focus is on \textit{iid} data, with non-\textit{iid} or sequential data appearing as an afterthought. The result of this is that there is a scarcity of automatic approaches to performing approximate Bayesian inference for generic dynamic, stochastic simulation models in the \gls{abc} literature with which practitioners of \gls{abc} can readily equip themselves. 
Developing automatic approaches to \gls{abc} that are more tailored to simulators generating sequences of dependent points will thus increase the ease with which \gls{abc} methods can be deployed in a broader range of real-world inference settings. 

In response to this challenge, we present here two novel methods for performing \gls{abc} for time series models that bypass the difficult problem of manually constructing summary statistics for sequential data. Our approach leverages so-called \emph{path signatures}, a key object in the mathematics of rough path theory and the theory of controlled differential equations \citep[see e.g.][]{LyonsT.J2007Dedb, lyons2014rough}.
Signatures have been employed successfully in a variety of machine learning tasks, from hand-gesture recognition \citep{Li2017} to the early identification of Alzheimer's disease \citep{Moore2019}, and constitute a natural feature set for multivariate and even irregularly sampled sequential data \citep{salvi2020computing}.
We demonstrate that the path signature can be employed either directly as a summary statistic or in the context of a semi-automatic projection approach to construct powerful distance measures for time series data in \gls{abc}, and further that such approaches can recover more accurate posterior estimates than existing techniques.

\subsection{Likelihood-free inference background}\label{sec:abc}
\glsresetall

In this section, we will recapitulate some standard approaches to \gls{abc} with an emphasis on time series data.  
Let $\mcal{X}^n$ be the space of all length $n$ sequences taking values in $\mcal{X}$ and suppose we have time series data $\by = (\bd{y}_{t_1}, \bd{y}_{t_2}, \ldots, \bd{y}_{t_n}) \in \mcal{X}^n$, observed at real times $0 = t_1 < t_2 < \ldots < t_n = T$, and assumed to have been drawn from the generative model with density $p(\by \mid \bth)$ parameterised by $\bth = (\bth_1, \ldots, \bth_p) \in \bTh \subseteq \mathbb{R}^p$. Given a prior distribution $\pi(\bth)$ on $\bTh$, the central object in Bayesian inference is the posterior distribution
\begin{equation}\label{eq:Bayes}
    \pi(\bth \mid \by) \propto p(\by \mid \bth) \pi(\bth).
\end{equation}
For simulation models, the likelihood function $p(\by \mid \bth)$ is commonly intractable, in the sense that it cannot be evaluated point-wise, making infeasible standard Bayesian approaches to posterior inference such as \gls{mcmc}. 
    
In such scenarios, an established alternative is offered by \gls{abc} \citep{tavare1997inferring, pritchard1999population, beaumont2002approximate} which allows the user to approximate the true posterior \eqref{eq:Bayes} using only forward samples from the simulator. Broadly, the user is required to specify summary statistics $\bs : \mcal{X}^n \to \mathbb{R}^k$ and a distance measure $\rho$, and the true likelihood function is approximated as
\begin{equation}\label{eq:ABC_likelih}
    \tilde{p}\left\{\bs\left(\by\right) \mid \bth\right\} = \int K_{\varepsilon}\left[\rho\{ \bs\left(\by \right), \bs\left(\bx\right)\} \right] \cdot p(\bx \mid \bth)\, \mathrm{d}\bx,
\end{equation}
where $K_{\varepsilon}(\cdot) = K(\cdot/\varepsilon)/\varepsilon$ is a kernel function with bandwidth parameter $\varepsilon$.
The resulting \gls{abc} posterior is then given by
\begin{equation}\label{eq:abc_post}
    \pi_{\gls{abc}}\left\{\bth \mid \bs\left(\by\right) \right\} \propto \tilde{p}\left\{\bs\left(\by\right) \mid \bth \right\} \pi\left(\bth \right),
\end{equation}
which is consistent as $\varepsilon\to 0$ if the employed summary statistic is sufficient, since
as $\varepsilon \rightarrow 0$, $K_{\varepsilon}\left[\rho\{\bs\left(\by\right), \bs\left(\bx\right)\}\right] \rightarrow \delta_{\by}(\bx)$, and so the right hand side of Equation \eqref{eq:ABC_likelih} approaches $p(\by \mid \bth)$.
Additionally, extending upon the concept of generalized Bayesian inference \citep{bissiri2016general, knoblauch2019generalized}, \citet{Schmon2020} note that \gls{abc} can be seen as a generalized Bayesian method targeting the posterior
\begin{equation}\label{eq:gbi}
    \pi_{\acrshort{gbi}}(\bth \mid \by) \propto \int e^{-w\cdot \ell(\by; \bx)} p(\bx \mid \bth) \pi(\bth) \d\bx
\end{equation}
for an arbitrary loss function $\ell(\by;\bx)$ that captures the discrepancy between observation $\by$ and simulation $\bx$, and some weight hyperparameter $w \in \mathbb{R}$.

The approach as presented above leaves open a plethora of possible choices for $\bs$, $\rho$ and $K_{\varepsilon}(\cdot)$---or, more generally, the loss function $\ell$---which has sparked great interest in the choice of those values in different scenarios, the complete enumeration of which is beyond the scope of this %
overview. However, we summarise here some of the most common approaches. 
\vspace{-0.3cm}
\paragraph{Rejection ABC}
The standard \gls{rej} algorithm corresponds to choosing a uniform kernel $K_{\varepsilon}(\cdot) \propto \mathbbm{1}\left(\cdot \leq \varepsilon\right)$. That is, parameter values $\bth$ are independently drawn from the prior and are retained as samples from an approximate posterior according to whether the distance between $\bs(\by)$ and $\bs(\bx)$ falls at or below a threshold $\varepsilon$. The choice of threshold $\varepsilon$ is left to the experimenter, and for example may be determined in advance of the inference procedure, or chosen after simulation time such that a certain proportion of the total simulation budget is retained.

\vspace{-0.3cm}
\paragraph{Semi-automatic ABC} \cite{Fearnhead2012} propose a method for automatically generating low-dimensional summary statistics by reducing a larger candidate set of summaries, referred to as \gls{saabc}. Their approach may be summarised as follows: Given a set of $N$ training data points $\left(\bx^{(i)}, \bth^{(i)}\right) \sim p( \bx, \bth)$, $i=1, \dots, N$, and a candidate vector $\mathbf{g}(\cdot)$ of $J$ summary statistics, the method proceeds by performing vector-valued linear regression from $\mathbf{g}(\bx^{(i)})$ to $\bth^{(i)}$, producing a matrix $A$ of coefficients. The summaries $\bs$ are then taken to be the output of this regression, i.e. $\bs(\bx^{(i)}) = A \mathbf{g}(\bx^{(i)})$. The motivation for this is that, under a quadratic loss, the optimal summary statistics can be shown to be the posterior mean $\mathbb{E}\left(\bth \mid \by\right)$. A drawback of this method, however, is that it requires the construction of an initial set of candidate summaries, which would need to be informative. %
Other approaches in this vein include that of \citet{Nakagome2013}, in which the authors propose the use of \gls{saabc} using kernel ridge regression, to exploit the nonlinearities induced by kernel methods in this regression task. 

\vspace{-0.3cm}
\paragraph{K2-ABC} \citet{Park2016} propose \gls{k2abc}, an \gls{abc} method that bypasses the problem of constructing summary statistics for \textit{iid} data by using the \gls{mmd} between (a) the simulator's distribution $f( \cdot \mid \bth)$, where $\bx = (\bx_1, \dots, \bx_n) \sim p(\bx \mid \bth) = \prod_{i=1}^{n} f(\bx_i \mid \bth)$, and (b) the true density $f^*$ giving rise to the \textit{iid} observations comprising $\by$, respectively. That is, with a suitable kernel $k$, the discrepancy between the simulation output $\bx$ and observation $\by$ is then taken to be the squared \gls{mmd}
\begin{equation}
    \textsc{mmd}^2 = \|\mathbb{E}_{\bd{z} \sim f(\cdot \mid \bth)}[k(\bd{z}, \cdot)] - \mathbb{E}_{\bd{z}' \sim f^*}[k(\bd{z}', \cdot)] \|_{\mathcal{H}}^2,
    \end{equation}
where $\mathcal{H}$ is the \gls{rkhs} associated with $k$. In this way, the choice of summary statistics (e.g. as required in \gls{saabc}) can be seen as being replaced by the choice of kernel $k$. For time series data, the authors suggest that the dependency structure can be ignored, and that the observation $\lbrace{\bd{y}_i : i=1, \dots, n\rbrace}$ and simulation output $\lbrace{\bd{x}_i : i = 1, \dots, m\rbrace} $ can still be treated as \textit{iid} data from the marginal densities $f(\cdot \mid \bth) := f_{\bth}$ and $f^*$, respectively. An unbiased estimate of the $\gls{mmd}$, under this assumption, can thus be obtained as
\begin{align}\nonumber
    \widehat{\textsc{mmd}}^2\left(f_{\bth}, f^*\right) 
    &= 
    \frac{1}{m(m-1)} \sum_{i \neq j} k(\bd{x}_i, \bd{x}_j) 
    + 
    \frac{1}{n(n-1)} \sum_{i \neq j} k(\bd{y}_i, \bd{y}_j)
    &- 
    \frac{2}{n m} \sum_{\substack{i = 1, \dots, m\\j = 1, \dots, n}} k(\bd{x}_i, \bd{y}_j).
\end{align}

\vspace{-0.8cm}
\paragraph{Wasserstein ABC} \citet{Bernton2019} propose a further method for measuring the discrepancy between observations and simulated data that circumvents the problem of manually constructing summary statistics. The approach uses as its measure of discrepancy the $p$-Wasserstein distance between the empirical distribution of observations $\by = (\bd{y}_1, \bd{y}_2, \dots, \bd{y}_n)$, and simulated data $\bx = (\bd{x}_1, \bd{x}_2, \dots, \bd{x}_m)$, with $\bd{y}_i, \bd{x}_j \in \mathbb{R}^d$. That is, the distance $\rho$ is taken to be
\begin{equation}\label{eq:wass}
    \mathcal{W}_p(\by, \bx)^p = \inf_{\gamma \in \Gamma_{n,m}} \sum_{i =1}^{n} \sum_{j=1}^{m} \rho_0(\bd{y}_i, \bd{x}_j)^p \gamma_{ij}
\end{equation}
where $\rho_0$ is a distance on $\mathbb{R}$ and $\Gamma_{n,m}$ is the set of $n\times m$ matrices with non-negative entries, columns summing to $m^{-1}$, and rows summing to $n^{-1}$. The authors propose to use $p = 1$, in order to make a minimal number of assumptions on the existence of moments of the data-generating process.

A number of solutions are proposed in \citet{Bernton2019} to account for the dependency structure inherent in time series data. The first strategy discussed is the use of \emph{curve matching}, in which a time augmentation $\bd{y}_{t_i} \mapsto (t_i, \bd{y}_{t_i})$ is applied to the data, and the following ground distance between elements of the sequence used:
\begin{equation}
    \rho_{0}\{(t_i, \bd{y}_{t_i}), (t_j, \bd{x}_{t_j}); \lambda\} = \| \bd{y}_{t_i} - \bd{x}_{t_j} \| + \lambda \vert{ t_i - t_j \vert}
\end{equation}
where $\lambda > 0$ is a free parameter that interpolates the distance in \eqref{eq:wass} between the sum of Euclidean distances $\sum_{i} \| \bd{y}_{t_i} - \bd{x}_{t_i} \|$ and the Wasserstein distance between the empirical marginal distributions of $\by$ and $\bx$. A heuristic for tuning $\lambda$ is offered only for the case of univariate $\by$ and $\bx$.%

A second strategy employs \emph{reconstructions}, where the data are transformed to generate empirical distributions that allow for easier identifiability of parameters. Two types of reconstructions are considered: delay reconstructions, which is a common technique for reconstructing phase spaces in dynamical systems theory that involves considering lagged sequences of observations from the data; and residual reconstructions, in which the data is transformed according to the structure of the generative model such that they become \emph{iid} observations, for example by considering $\epsilon_t = \left(\bd{x}_t - a\, \bd{x}_{t-1}\right)/\sigma$ in the case of a centered AR(1) model with parameter $\bth = (a, \sigma)$. However, it is undesirable to rely on such methods. For the case of delay reconstructions, properly estimating the lag parameters is key to its success \citep{PhysRevA.33.1134} and obtaining reliable estimates remains a significant challenge \citep{Bradley_2015}. This is likely to be exacerbated in \gls{lfi} settings, in which time series are stochastic and are often short, due to computational expense. Delay reconstructions will then also further reduce this length of the data, which can be costly to the quality of the inference procedure. Furthermore, the often complicated or unknown internal mechanisms of complex simulation models typically do not allow for a simple transformation of the output into \emph{iid} data, limiting the applicability of this approach in \gls{lfi} settings.

\vspace{-0.3cm}
\section{Path signatures}\label{sec:Back}
There are currently few methods well-suited to performing approximate Bayesian inference for general time series models. %
Existing approaches often make restrictive assumptions such as \emph{iid} data, or require the use of data transformations that are difficult to construct or that involve a potentially substantial reduction in the length of datasets, which may be prohibitively costly.
Moreover, where solutions for time series are proposed, their discussion is often limited to univariate data, raising the question of whether methods exist that are robust to more general \gls{lfi} settings, such as those involving multivariate and/or irregularly sampled data with missing values or data evolving on general topological spaces.
To this end, we introduce the use of \emph{path signatures} as a flexible and general framework for performing \gls{lfi} for complex time series models, and provide an overview of their important properties in this section. 

Let $\mcal{H}$ be a Hilbert space and $h : [0,T] \to \mcal{H}$ be a $\mcal{H}$-valued path on interval $[0,T]$. For $p \geq 1$, we denote the $p$-variation of $h$ over the interval $[s,t] \subseteq [0,T]$ as
\begin{equation*}
    \norm{h}{p-\text{var},[s,t]} := \left(\sup_{\zeta(s,t)} \sum_{i=1}^{n - 1} \norm{h_{t_{i+1}} - h_{t_i}}{\mcal{H}}^p\right)^{1/p}
\end{equation*}
where the supremum is taken over all finite partitions $\zeta(s,t)$ of the domain and $n = \vert{\zeta(s,t)}\vert$. 
Throughout this work, we will primarily consider $\mcal{H}$-valued paths of bounded variation over the entire interval $[0,T]$, i.e. paths of finite $p$-variation for $p=1$ such that
\begin{equation*}
    \onevar{h} := \sup_{\zeta(0,T)} \sum_{i=1}^{n - 1} \norm{h_{t_{i+1}} - h_{t_i}}{\mcal{H}} < \infty,
\end{equation*}
where the interval $[0,T]$ is omitted from the subscript for simplicity. We denote with $\bv{\mcal{H}}$ the space of all such paths. The \emph{path signature} \citep[see e.g.][]{LyonsT.J2007Dedb} of $h$, denoted $\sig{h}$, maps such paths to an infinite series of tensors:
\begin{equation}\label{eq:sig}
    \text{Sig} : \bv{\mcal{H}} \to \pH,\ \ \ h \mapsto \{1, S_1(h), S_2(h), \dots\},
\end{equation}
where
\begin{equation}
    \pH := \mathbb{R} \oplus \mcal{H} \oplus \left(\mcal{H} \otimes \mcal{H}\right) \oplus \dots \oplus \mcal{H}^{\otimes m} \oplus \dots
\end{equation}
and where we define recursively
\begin{equation}\label{eq:sig_terms}
    S_m := \int_{0}^T {\d }h^{\otimes m} := \int_{0}^T \int_{0}^t {\d }h^{\otimes (m-1)} \otimes {\d }h_t.
\end{equation}
In the above, we have adopted the convention that $\mcal{H}^{\otimes 0} = \mathbb{R}$. We expand on this introduction of signatures for the unfamiliar reader in Section \ref{app:signatures} of the appendix.

\subsection{Key properties of path signatures}
Signatures have a number of desirable properties. In the following subsections, we consider some of the main properties that we will make use of throughout this work.
\vspace{-0.5cm}
\subsubsection{Universal nonlinearity}
\vspace{-0.3cm}
One such property is \emph{universal nonlinearity}: the signature captures all possible nonlinearities in path-valued random variables, in the sense that it is possible to approximate any nonlinear function of a path arbitrarily well with a linear functional of the signature. This is a consequence of the \emph{shuffle product} property of signatures (see Section \ref{app:shuffle_sig} of the appendix). 
Applying the classical Stone-Weierstrass theorem\footnote{An issue that arises in the application of the classical Stone-Weierstrass theorem in this context is that the space of interest to us -- $\bv{\mcal{H}}$ -- is not locally compact. The classical Stone-Weierstrass theorem therefore cannot strictly be applied here. However, \citet{chevyrev2018signature} demonstrate that a Stone-Weierstrass result exists by equipping the space of continuous bounded real-valued functions on $\bv{\mcal{H}}$ with an appropriate topology. See \citet{chevyrev2018signature} for details.} results in the stated universal nonlinearity property, which can be formalised as follows:
\vspace{-0.2cm}
\begin{Theorem}
    Let $\mcal{K}$ be a compact set of non-tree-like\footnote{See Section \ref{sec:invariances}.} paths of bounded variation, and $C(\mcal{K}, \mathbb{R})$ be the space of continuous, real-valued function on $\mcal{K}$. Then the space of linear functionals on signatures of paths in $\mcal{K}$ is dense in $C(\mcal{K}, \mathbb{R})$; that is, for any $f \in C(\mcal{K}, \mathbb{R})$ and any $\varepsilon>0$, there exists an $L \in \bigoplus_{m\geq 0}\mcal{H}^{\otimes m}$ such that
    \begin{equation*}
        \sup_{h\in \mathcal{K}}\Big|f(h) - L\{\sig{h}\}\Big| < \varepsilon.
    \end{equation*}    
\end{Theorem}
\vspace{-0.8cm}
\subsubsection{Invariance properties}\label{sec:invariances}
\vspace{-0.3cm}
Further properties of the signature include its translation and reparameterisation invariance:
\vspace{-0.5cm}
\begin{Proposition}
    Let $h \in \bv{\mcal{H}}$, $a \in \mcal{H}$, and $\psi : [0,T] \to [0,T]$. Then $\sig{h + a} = \sig{h}$ and $\sig{h \circ \psi} = \sig{h}$. 
\end{Proposition}
\vspace{-0.1cm}
In this way, signatures are able to factor out nuisance and potentially infinite-dimensional symmetries where this is beneficial. However, when such invariances are disadvantageous, they can easily be destroyed with two extremely simple preprocessing techniques: \textit{time-augmentation}, in which the path $(t, h_t)$ is instead considered, and \textit{basepoint augmentation}, in which $h_0 = c$ for some fixed constant $c \in \mcal{H}$ is enforced for all paths under consideration.

A third, more interesting invariance property results from the signature's inability to identify regions of the path in which, informally speaking, a retracing of the path occurs \citep{chen1958integration, Hambly_2010, BOEDIHARDJO2016720}; that is, for example, paths of the form $a \star b \star \overleftarrow{b} \star c$ for $a, b, c \in \bv{\mcal{H}}$, where $\star$ denotes concatenation and $\overleftarrow{b}$ is the path $b$ ``run-backwards''. Paths in which such retracings occur are referred to as \emph{tree-like equivalent} to their reduced paths such that, for example, $a \star b \star \overleftarrow{b} \star c \sim_t a \star c$, where $\sim_t$ denotes tree-like equivalence. 
While this phenomenom has previously been studied in more specific cases \citep{chen1958integration, Hambly_2010}, the most general form of this invariance property is provided by \citet{BOEDIHARDJO2016720}, a special case of which may be stated as follows:
\vspace{-0.2cm}
\begin{Theorem}[\citet{BOEDIHARDJO2016720}]\label{thm:sig_inj_tle}
    Let $V$ be a Banach space and $h, g \in \bv{V}$. Then $\sig{h} = \sig{g}$ iff $h \sim_t g$.
\end{Theorem}
\vspace{-0.2cm}
In the real world, however, tree-like equivalent paths are rare and can straightforwardly be avoided by considering only time-augmented paths $h : [0,T] \to \mcal{H} \times [0,T],\ t \mapsto (t, h_t)$. Such a transformation ensures that the path is injective, meaning no partial retracing can occur at any point along the path. This, along with their universal nonlinearity property, demonstrates that signatures are powerful and faithful representations of paths and are, essentially, an injective feature map for path-valued random variables. Signatures are therefore an appealing option for performing inference for dynamic, stochastic processes.

\subsection{The signature kernel}

Computing iterated integrals for high- or potentially infinite-dimensional paths quickly becomes computationally infeasible due to the combinatorial explosion of terms in the signature with increasing depth. In part due to this, recent research effort \citep{Kiraly2019, salvi2020computing} has been directed towards kernelising the feature map in Equation \eqref{eq:sig}, permitting the use of the signature in learning procedures without explicit evaluation of the signature terms themselves. We provide here further details on the resultant \emph{signature kernel}, of which we make use throughout the current work.

We follow \citet{Kiraly2019} and begin by defining the following for $A, B \in \pH$:
\begin{equation}
    A + B := (a_0 + b_0, a_1 + b_1, \dots)
\end{equation}
and an inner product
\begin{equation}\label{eq:inn_prod_ta}
    \ip{A}{B}{} := \sum_{m \geq 0} \ip{a_m}{b_m}{\mcal{H}^{\otimes m}},
\end{equation}
where $A = (a_0, a_1, \dots)$, $B = (b_0, b_1, \dots)$, and
\begin{equation}
    \ip{u_1 \otimes \dots \otimes u_m}{v_1 \otimes \dots \otimes v_m}{\mcal{H}^{\otimes m}} = \prod_{j=1}^m \ip{u_j}{v_j}{\mcal{H}}.
\end{equation}
This leads us to the following norm on $\pH$:
\begin{equation}
    \norm{A}{} := \sqrt{\sum_{m \geq 0} \norm{a_m}{\mcal{H}^{\otimes m}}^2}.
\end{equation}
Using the inner product \eqref{eq:inn_prod_ta} and the fact that $\sig{h} \in \pH$ for $h \in\bv{\mcal{H}}$, we arrive at the definition of the signature kernel:

\begin{Definition}[Signature kernel, \citet{Kiraly2019}]\label{def:sig_kern}
    The \emph{signature kernel} for $h, g \in \bv{\mcal{H}}$ is
    \begin{equation}\label{eq:sig_kern}
        k : \bv{\mcal{H}} \times \bv{\mcal{H}} \to \mathbb{R}, \ \ \ (h, g) \mapsto \ip{\sig{h}}{\sig{g}}{},
    \end{equation}
    where the inner product is defined as in Equation \eqref{eq:inn_prod_ta}.
\end{Definition}
A key insight of \citet{Kiraly2019} was to recognise that evaluation of the signature kernel -- which operates on \textit{paths} in $\mcal{H}$ -- can be performed using only evaluations of an inner product $\kappa$ that operates on \textit{points} in the path, amounting to a kernel trick for the signature kernel. \citet{Kiraly2019} further describe an efficient Horner scheme to evaluate a truncated signature kernel that approximates Equation \eqref{eq:sig_kern}. In more recent work, \citet{salvi2020computing} provide an alternative approach to approximating Equation \eqref{eq:sig_kern} without truncation by observing that the signature kernel solves a Goursat partial differential equation. The solution to this Goursat problem may be obtained numerically with standard finite element methods, and can similarly be computed using only evaluations of an inner product $\kappa$ on points in the path.

\subsection{Path signatures in practice}

In light of their interesting and useful properties described above, signatures can be seen as a canonical feature transformation for path-valued random variables. 
However, there exists an incongruity between our discussion so far and the scenarios faced in real-world settings: in reality and from the output of simulation models, we tend to observe discretely sampled data $\bd{x} = (\bd{x}_{t_1}, \bd{x}_{t_2}, \dots, \bd{x}_{t_n})$ at times $0 = t_1 < t_2 < \dots < t_n = T$, where $\bd{x}_{t} \in \mcal{X}$ for some finite-dimensional space $\mcal{X}$ (for example $\mathbb{R}^d$ or $\mathbb{R}^{d \times d}$ for some $d \geq 1$), rather than continuous paths $x \in \bv{\mcal{H}}$. 
This is dealt with naturally in the signature (kernel) literature in the following ways: 
\begin{enumerate}
    \item[(a)] As noted by \citet{Kiraly2019}, the aforementioned signature kernel trick can be used to introduce nonlinearities and embed the $\mcal{X}$-valued sequence $\bd{x}$ in a Hilbert space. In particular, by choosing a reproducing kernel $\kappa : \mcal{X} \times \mcal{X} \to \mathbb{R}$ with \gls{rkhs} $\mcal{H}$ and canonical feature map $\kap{\bd{x}_{t}} \in \mcal{H}$ as the inner product on the data space $\mcal{X}$, we may implicitly construct a sequence $(\kap{\bd{x}_{t_1}}, \kap{\bd{x}_{t_2}}, \dots, \kap{\bd{x}_{t_n}})$ of points in $\mcal{H}$ from sequences of data in $\mcal{X}$.
    \item[(b)] To construct continuous paths from the discrete sequence above, an interpolation scheme is employed. While many interpolation schemes are possible, the most common is linear interpolation. Indeed, \citet{Kiraly2019} and \citet{salvi2020computing} assume a linear interpolation to construct \textit{discretised} signature kernels operating on sequences of points, and we use this interpolation scheme throughout this work.
\end{enumerate}
By combining the above two steps, we may progress from a sequence $\bd{x}$ of points in $\mcal{X}$ to a $\mcal{H}$-valued, piecewise linear path $h$, given by
\begin{equation}\label{eq:lifted_path}
    h_t := \kap{\bd{x}_{t_{i}}} + \frac{t - t_{i}}{t_{i+1} - t_{i}} \{\kap{\bd{x}_{t_{i+1}}} - \kap{\bd{x}_{t_{i}}}\} \text{ for } t \in \left[t_{i}, t_{i+1}\right],\ i = 1, \dots, n-1.
\end{equation}
Piecewise linear paths constructed in this way are naturally of bounded variation if, for example, $\kappa$ is a continuous and/or 
uniformly bounded kernel\footnote{See Proposition \ref{prop:sig_inj_abc} below.}. We will assume this throughout, such that all observed sequences in $\mcal{X}$ lift to piecewise linear paths of bounded variation in $\mcal{H}$ under the feature map corresponding to $\kappa$, and denote the space of piecewise linear paths of bounded variation in $\mcal{H}$ over time interval $[0,T]$ with $\Pl{\mcal{H}}$. We will furthermore abuse notation slightly by letting $\kap{\bd{x}} \in \Pl{\mcal{H}}$ denote the path in Equation \eqref{eq:lifted_path}, i.e. the linear interpolation of the lifted points $(\kap{\bd{x}_{t_1}}, \kap{\bd{x}_{t_2}}, \dots, \kap{\bd{x}_{t_n}})$, while denoting the feature map for $\bd{x}_{t}$ with $\kap{\bd{x}_t} \in \mcal{H}$. Finally, we will take $k(\bd{x}, \cdot) := \sig{\bd{x}}$ to mean the signature of the piecewise linear, $\mcal{H}$-valued path $\kap{\bd{x}}$, while $\sig{g}$ denotes the signature of a path $g \in \bv{\mcal{H}}$.

\vspace{-0.5cm}
\subsubsection{Further pre-processing}\label{sec:PathTrans}
\vspace{-0.2cm}
Prior to lifting the sequence to a path in $\mcal{H}$, and depending on the nature of the data at hand, it is sometimes appropriate to apply a transformation to the data: %
certain transformations may enable the signature to represent information in the stream more conveniently for the learning task at hand. A large set of such transformations have been proposed in the literature on inference using path signatures; see \citet{Morrill2020} for a recent summary and comparison of many of these. Here, we describe two such pre-signature transformations that we will use in this paper. %

\vspace{-0.5cm}
\paragraph{Cumulative sum} Recall from Figure \ref{fig:geom} that the depth 1 signature terms correspond to the increment along the path, and that a subset of the depth 2 terms correspond to the areas above and below the curve. For certain data types, for example non-negative binary or spiking data, the data may not be well-characterised by these terms by default. In such cases it can be beneficial to consider instead the cumulative sum of the observations \citep{Kiraly2019}, which can intuitively be thought of as propagating information from earlier in the sequence to later in the stream, more readily exhibiting the structure of the stream. The effect of this can be to shift information into lower order terms in the signature, for example the increments (depth 1 terms).

\vspace{-0.4cm}
\paragraph{Delay transformation} A similar transformation to the above is a delay transformation, for example the lag-1 delay transformation:
\begin{equation}
    (\bd{x}_{t_1}, \bd{x}_{t_2}, \dots, \bd{x}_{t_n}) \mapsto ((\bd{x}_{t_1}, \bd{x}_{t_2}), (\bd{x}_{t_2}, \bd{x}_{t_3}), \dots, (\bd{x}_{t_{n-1}}, \bd{x}_{t_n})).
\end{equation}
While the number of channels doubles here also, this transformation may be computationally preferable to the lead-lag transformation, since the length of the sequence does not increase in this case.

\vspace{-0.3cm}
\subsubsection{Augmentations}
As noted previously, two augmentations can be applied to remove the signature's translation and reparameterisation invariance properties:

\textbf{Time augmentation}, in which the uniformly increasing time index $0 = t_1 < t_2 < \dots < t_n = T$ is added as a channel in the sequence:
\begin{equation}
    (\bd{x}_{t_1}, \bd{x}_{t_2}, \dots, \bd{x}_{t_n}) \mapsto \left((t_1, \bd{x}_{t_1}), (t_2, \bd{x}_{t_2}), \dots, (t_n, \bd{x}_{t_n})\right),
\end{equation}
denoting the times at which the points in the series occurred.

\textbf{Basepoint augmentation}, in which all sequences are enforced to assume a common but otherwise arbitrary initial value. This can be achieved by simply concatenating an arbitrary constant value to the beginning of each sequence.

\section{Approximate Bayesian computation with path signatures}\label{sec:signatureABC}

Given its unique properties, the path signature and its associated kernel
are natural candidates for feature maps and discrepancy measures in \gls{abc} to handle irregularly spaced and potentially multivariate time series data.
In this section, we will introduce and investigate two simple but powerful techniques for incorporating signatures in \gls{abc}.

\subsection{Signature ABC}\label{sec:skd}

\glsreset{rej}

Though signatures are infinite-dimensional objects, we can leverage their kernel representation (see Definition \ref{def:sig_kern}) to compute the distance between two sequences $\bd{x}, \bd{y}$ as the norm induced by the associated signature inner product. That is, for two time series $\bd{x}$ and $\bd{y}$, we can interpret the signature of their lifted paths as a \emph{summary statistic}, $\bd{s}(\bd{x}) = \sig{\bd{x}}$, and compute
\begin{equation}
\label{eq:sig_distance}
    \rho\{\bd{s}(\bd{x}), \bd{s}(\bd{y})\} := \| \sig{\bd{x}} - \sig{\bd{y}} \|^2 = k(\bd{x}, \bd{x}) + k(\bd{y}, \bd{y}) - 2\,k(\bd{x}, \bd{y}),
\end{equation}
where, again, $k(\bd{x}, \bd{y}) = \ip{\sig{\bd{x}}}{\sig{\bd{y}}}{}$.
The resulting distance can be computed easily using, for example, the \texttt{sigkernel}\footnote{\url{https://github.com/crispitagorico/sigkernel}} package (see \ref{app:sabc_code} for an example implementation) or alternatives\footnote{See e.g. \url{https://github.com/tgcsaba/KSig}.} and used to derive an \gls{abc} posterior via Equations \eqref{eq:ABC_likelih}-\eqref{eq:abc_post}. For example, it may be embedded either in rejection \gls{abc}, leading to the \gls{abc} posterior
\begin{equation*}
    \pi_{\textsc{rej}}(\bth \mid \bd{y}) \propto \pi(\bth) \int 1\left(\| \sig{\bd{x}} - \sig{\bd{y}} \|^2 \leq \varepsilon\right) p(\bd{x} \mid \bth) {\d } \bd{x},
\end{equation*}
or alternatively following the approach of \citet{Schmon2020} as a loss in the generalized approximate posterior \eqref{eq:gbi}, that is
\begin{equation*}
    \pi_{\textsc{gbi}}(\bth \mid \bd{y}) \propto \pi(\bth) \int e^{- w \| \sig{\bd{x}} - \sig{\bd{y}} \|^2 } p(\bd{x} \mid \bth) {\d } \bd{x}.
\end{equation*}
In both cases our method straightforwardly extends classical approaches by using the distance function \eqref{eq:sig_distance}, suggesting the name \gls{sabc}. 
In the latter case, Monte Carlo samples can be obtained using, for example, a pseudo-marginal approach \citep{beaumont2003estimation, andrieu2009pseudo}. For the remainder of this paper, however, we will only consider standard \gls{rej} in the interest of a simple and fair comparison with alternative distance measures. 

We next consider the theoretical properties of the \gls{sabc} posterior. In particular, we consider two asymptotic regimes: the correctness of the \gls{sabc} posterior for fixed data and as the \gls{abc} tolerance hyperparameter $\varepsilon \to 0$; and the behaviour of the \gls{sabc} posterior for fixed $\varepsilon$ and as the number of samples $n\to \infty$ in the interval $[0,T]$ or, equivalently, as the sampling rate tends to infinity. 

\subsubsection{Behaviour as $\varepsilon \to 0$ for fixed $n$}

We first demonstrate that the discrepancy measure in Equation \eqref{eq:sig_distance} satisfies the conditions specified in Proposition 3.1 of \citet{Bernton2019}, which gives a statement on the convergence of \gls{abc} posteriors to the true posterior under certain regularity conditions on the simulator's likelihood function as $\varepsilon\to 0$. A specific case of the statement is as follows:
\begin{Proposition}[Proposition 3.1, \citet{Bernton2019}]\label{thm:abc_convergence}
    Let $\mcal{X} := \mathbb{R}^d$, $\bd{y} = (\bd{y}_1, \dots, \bd{y}_n) \in \mcal{X}^n$, and $\mcal{D} : \mcal{X}^n \times \mcal{X}^n \to \mathbb{R}_{\geq 0}$ be a non-negative distance measure on $\mcal{X}^n$. Suppose $p(\bd{x} \mid \bth)$ is the continuous density associated with simulated data $\bd{x} \in \mcal{X}^n$ and that
    \begin{equation*}
        \sup_{\bth \in \bTh \setminus \mcal{N}_{\bTh}} p(\bd{x} \mid \bth) < \infty,
    \end{equation*}
    where $\mcal{N}_{\bTh}$ is a set such that $\pi(\bth) = 0\, \forall \bth \in \mcal{N}_{\bTh}$. Suppose further that there exists $\bar{\varepsilon} > 0$ such that
    \begin{equation*}
        \sup_{\bth \in \bTh \setminus \mcal{N}_{\bTh}} \sup_{\bd{z} \in \mcal{A}^{\bar{\varepsilon}}} p(\bd{z} \mid \bth) < \infty,
    \end{equation*}
    where $\mcal{A}^{\bar{\epsilon}} := \lbrace{\bd{z} : \mcal{D}(\bd{y}, \bd{z}) \leq \bar{\varepsilon} }\rbrace$. Suppose that $\mcal{D}$ is continuous. If $\mcal{D}(\bd{y}, \bd{z}) = 0$ iff $\bd{y} = \bd{z}$ then, keeping $\bd{y}$ fixed, the \gls{abc} posterior converges strongly to the posterior as $\varepsilon \to 0$.
\end{Proposition}

Therefore, provided that the stated regularity conditions on the simulator's likelihood function are met, showing that the distance function in Equation \eqref{eq:sig_distance} is continuous and injective is sufficient to show that the \gls{sabc} posterior converges to the true posterior as $\varepsilon \to 0$. These requirements are indeed met under the assumptions of Theorem \ref{thm:abc_convergence} and under additional benign conditions:
\begin{Proposition}\label{prop:sig_cont_abc}
    Let $\mcal{X} := \mathbb{R}^d$, $\bd{y} = (\bd{y}_1, \dots, \bd{y}_n) \in \mcal{X}^n$ be the fixed real-world dataset, and $\mcal{D}(\bd{y}, \cdot)$ be as in Equation \eqref{eq:sig_distance}, i.e.
    \begin{equation*}
        \mcal{D}(\bd{y}, \cdot) : \mcal{X}^n \to \mathbb{R}_{\geq 0},\ \ \ \bd{x} \mapsto \norm{\sig{\bd{y}} - \sig{\bd{x}}}{}^2. 
    \end{equation*}
    Assume both $\bd{y}$ and $\bd{x}$ are time- and basepoint-augmented, and that $\kappa : \mcal{X} \times \mcal{X} \to \mathbb{R}$ is a uniformly bounded kernel with continuous, injective canonical feature map. Then $\mcal{D}(\bd{y}, \cdot)$ is uniformly continuous.
\end{Proposition}
We defer the proof to Section \ref{app:sabc_proof_a} of the appendix. 
Injectivity of the signature map is also guaranteed under these conditions: 
\begin{Proposition}\label{prop:sig_inj_abc}
    Let $\mcal{X} := \mathbb{R}^d$, $\bd{x}, \bd{y} \in \mcal{X}^n$. Assume both $\bd{x}$ and $\bd{y}$ are time- and basepoint-augmented, and that $\kappa : \mcal{X} \times \mcal{X} \to \mathbb{R}$ is a uniformly bounded kernel with continuous, injective canonical feature map. Then $\sig{\bd{x}} = \sig{\bd{y}}$ iff $\bd{x} = \bd{y}$.
\end{Proposition}
The proof is given in Section \ref{app:sabc_proof_b} the appendix. 
Taken together, these results provide the same 
guarantees for 
the asymptotic correctness of the \gls{sabc} posterior as $\varepsilon\to 0$ for dynamic, stochastic simulators as, for example, the Wasserstein \gls{abc} posterior of \citet{Bernton2019}.

\subsubsection{Behaviour as $n \to \infty$ for fixed $\varepsilon$}

We now consider the behaviour of the \gls{sabc} posterior as the rate at which a (continuous) path is sampled tends to infinity, such that $n \to \infty$ within a fixed, finite time interval $[0,T]$. For the moment, we will assume that the continuous $\mcal{H}$-valued paths $h,g$ of which $\kap{\bd{x}}$ and $\kap{\bd{z}}$ are discretisations are of bounded variation, and will discuss a more general setting later. From this, we have that the \gls{sabc} posterior for piecewise linear paths converges to the \gls{sabc} posterior for continuous paths of bounded variation as the sampling rate is increased indefinitely:

\begin{Proposition}\label{prop:mesh_prop}
    Let $\kappa$ be a uniformly bounded, injective kernel and $g \in \bv{\mcal{H}}$ be the limit of $\kap{\bd{y}} \in \Pl{\mcal{H}}$ as $\mesh\to 0$. Then for fixed $\varepsilon > 0$ such that 
    \begin{equation*}
        \varepsilon > \inf_{h' \in \bv{\mcal{H}}} \mcal{D}(h', g)
    \end{equation*}
    and as $n \to \infty$ (equiv. $\mesh \to 0$), the \gls{sabc} posterior
    \begin{equation*}
        \pi\{\bth \mid \mcal{D}(\bd{x}, \bd{y}) \leq \varepsilon \} \rightharpoonup \pi\{\bth \mid \mcal{D}(h, g) \leq \varepsilon \}
    \end{equation*}
    for $h,g \in \bv{\mcal{H}}$, where $\rightharpoonup$ denotes weak convergence.
\end{Proposition}

We provide the proof in Section \ref{app:mesh_proof} the appendix. 
This result shows that for fixed $\varepsilon$ greater than the minimum possible value for $\mcal{D}(h, g)$, the \gls{sabc} posterior does not converge to a Dirac mass in the limit of infinite data over a fixed finite time horizon, or as the sampling rate is increased indefinitely in the interval $[0,T]$. Furthermore, by the same reasoning as in \citet[][Theorem 5.6]{miller2018robust}, continuity of the signature in the 1-variation topology (see Section \ref{prop:sig_cont_abc} of the appendix 
and \citet[][Section 3.1.2]{lyons2002system}) implies that the \gls{sabc} posterior is robust to small changes in the data even in the limit of infinite data. As the authors discuss, this can be advantageous in misspecified settings, which is typically the case in real-world modelling and inference problems.

\begin{Remark}
    Throughout the above, we have assumed that the limiting paths are of bounded variation as $\mesh \to 0$. 
    We may consider a more general case by adopting the weaker assumption that the limiting paths $h$ and $g$ for $\kap{\bx}$ and $\kap{\bd{y}}$ as $\mesh \to 0$ are \emph{geometric} $p$-rough paths (see Section \ref{app:rough_paths} of the appendix). By the Extension Theorem (see Section \ref{app:rough_paths} of the appendix and \citet[][Theorem 3.1.3]{lyons2002system}), the iterated integrals comprising the geometric $p$-rough paths $h$ and $g$ may be extended to all iterated integrals to obtain a path signature for $h$ and $g$ that is continuous in the $p$-variation topology (see \citet[][Theorem 3.10]{LyonsT.J2007Dedb} and Section \ref{app:rough_paths} of the appendix). In this way, we may obtain \gls{sabc} posteriors in the limit $\mesh \to 0$ for classes of models that are much ``rougher'' than the bounded variation case considered so far, such as continuous semimartingales, Gaussian processes, continuous-time Markov processes etc. The ``coarsened'' posteriors (using the nomenclature introduced by \citet{miller2018robust}) resulting from the application of \gls{sabc} in these instances are equipped with the same continuity property, now in the $p$-variation topology, that the \gls{sabc} posterior enjoyed in the bounded variation case under the 1-variation topology.
\end{Remark}

\subsection{Signature Regression ABC}

In some circumstances, it is desirable to find low-dimensional summary statistics for use in \gls{abc}. For example, \citet{Fearnhead2012} propose the use of the posterior mean $\mathbb{E}\left(\bth \mid \by \right)$ as a summary statistic for $\by$, since it is an optimal choice in that it minimises the quadratic loss between the \gls{abc} posterior mean and the true parameter. As discussed in Section \ref{sec:abc}, this involves fitting a vector-valued regression model from a large candidate set of summary statistics to parameters $\bth$, since this generates an estimate of the (unknown) posterior mean. The approach of \citet{Fearnhead2012} belongs to a larger class of methods for generating low-dimensional summary statistics from a large initial candidate set, sometimes %
termed ``projection methods'' \citep{beaumont2019approximate}, which also includes the partial least regression method proposed by \citet{10.1534/genetics.109.102509}.

However, a significant problem with projection methods is that it is often unclear which summary statistics should be included in the initial candidate set. Yet, the efficacy of the approach requires this initial candidate set to contain informative summaries in the first place. Contriving informative statistics thus represents a major obstacle in many inference tasks, and can involve significant domain expertise, experimentation, and computational expense. Consequently, when low-dimensional summary statistics are desired, it would be preferable to bypass the manual construction of an initial candidate set of statistics in order to use projection methods.

For the case of time series models, the path signature is a natural set of summary statistics for the regression task in \gls{saabc}, providing a basis for learning functions on streams due to its unique universal nonlinearity property. Naive regression on the full path signature is of course impossible, since the signature is an infinite-dimensional object. However, this may once again be circumvented using the signature kernel and corresponding kernel trick (see Definition \ref{def:sig_kern}), in the following way: use the signature kernel and kernel ridge regression \citep{HastieTrevor2001Teos} to implicitly regress parameters onto the \emph{full} signature, which is in a sense equivalent to using the infinitely long path signature as the candidate set of summary statistics in semi-automatic \gls{abc}. That is, using training examples $\lbrace{\bx^{(i)}, \bth^{(i)}\rbrace}_{i=1}^{R} \sim p\left(\bx \mid \bth\right) \pi\left(\bth\right)$, we find a function $\hat{\bth}_j$ in the \gls{rkhs} associated with the signature kernel $k$, which by the Representer Theorem has the following form for each component $\bth_j, j = 1, \dots, p$ of the $p$-dimensional parameters $\lbrace{\bth^{(i)}\rbrace}_{i=1}^{R}$:
\begin{equation*}
    \hat\bth_{j}(\bx) = \sum_{i=1}^{R} \boldsymbol{\omega}_i^{(j)} k(\bx, \bx^{(i)})
\end{equation*}
with
\begin{gather*}
    \boldsymbol{\omega}^{(j)} = \left(G + \alpha I_R\right)^{-1}\bpsi^{(j)},\quad \quad \quad G_{mn} = k(\bx^{(m)}, \bx^{(n)}),\\[1ex] %
    \bpsi^{(j)} = 
        \left[
        \begin{array}{c}
            \bth_{j}^{(1)}\\ \bth_{j}^{(2)}\\ \vdots\\ \bth_{j}^{(R)}
        \end{array}
        \right],
        \quad \quad I_R = \text{diag}(1, 1, \dots, 1) \in \mathbb{R}^{R\times R},
\end{gather*}
and 
$\alpha \geq 0$ is a regularisation parameter to be tuned. In this sense, signatures not only provide a natural notion of distance between time series, as described in Section \ref{sec:skd}, but additionally provide a suitable basis for learning functions on sequences, enabling the semi-automatic construction of summary statistics. This approach to \gls{abc} is somewhat similar to that of \citet{Nakagome2013}, who employ kernel ridge regression with a Gaussian RBF kernel to perform \gls{saabc}. Our approach differs substantially, however, in that \citet{Nakagome2013} propose the use of hand-crafted summary statistics as input to the kernel ridge regression model, while we propose the use of the full data. 

Once the data is summarised using this regression model, the discrepancy between simulation and observation is then computed as the Euclidean distance between their corresponding outputs from the kernel ridge regression model. We herein refer to this approach as \gls{skrr}, and provide further mathematical details on this approach in Section \ref{app:skabc} of the appendix.

\subsection{Computational complexity}

Evaluating the signature kernel for two streams $\by \in \mcal{X}^n$ and $\bx \in \mcal{X}^m$ with $\mcal{X} = \mathbb{R}^d$ has complexity that is linear in $d$ and linear in the product $nm$ \citep{salvi2020computing}. 
This is likewise the case for \gls{mmd}, which has complexity $\mathcal{O}\left(n^2\right)$ \citep{Park2016}, and compares favourably with 
\gls{wass}, which in multivariate settings is known to scale poorly with the number of data. \citet{Bernton2019}, for example, note costs of order $n^3$ when the Hungarian algorithm is used to solve the assignment problem. Alternative algorithms with favourable performance (compared to the Hungarian algorithm) are an active area of research, however scalability with data remains a problem for the application of Wasserstein \gls{abc} in large data settings.

\section{Experiments}\label{sec:Exp}

In this section, we present experiments comparing the performance of our signature-based methods against alternative notions of distance between simulation and observation. In particular, we compare our methods, signature \textsc{abc} (\gls{sabc}) and signature regression \textsc{abc} (\gls{skrr}), against the use of \gls{wass} \citep{Bernton2019} and \gls{mmd} \citep{Park2016} as measures of discrepancy, along with \gls{saabc} \citep{Fearnhead2012}. All code for reproducing these experiments is available on GitHub at \url{https://github.com/joelnmdyer/SignatureABC}.

\subsection{Implementation details}\label{sec:ExpImpDet}

For all losses, we sample from the \gls{abc} posterior using the simple rejection scheme outlined in Algorithm \ref{alg:Rej} and, unless stated otherwise, use $N=10^5$ and $M=10^3$. While other, more sophisticated schemes exist, we choose this to facilitate a simple and transparent comparison of the different distance measures. 
To assess the quality of the recovered posteriors, we compute the 1-Wasserstein distance and an unbiased estimate of the maximum mean discrepancy (MMD) between the approximate ground truth posteriors $\hat{\pi}_{\cdot \mid \by}$ and empirical posteriors $\hat{\pi}_{\mathrm{ABC}}$. In both cases, smaller values indicate a closer match to the approximate ground truth. To estimate the MMD between posteriors, we use a Gaussian RBF kernel with scale parameter chosen according to the median heuristic \citep{Briol2019}. All other implementation details are provided in Section \ref{app:imp_det} of the appendix.

\begin{algorithm}[t]
\SetAlgoLined
\textbf{Input:} prior $\pi$, observation $\by$, distance function $\mcal{D}(\cdot , \cdot)$, number of particles $N$, final sample size $M < N$\;
\KwResult{Empirical posterior $\sum_{i=1}^M \delta_{\bth^{(i)}}$}
 \For{$i=1,\dots,N$}{
  Sample $\bth^{(i)} \sim \pi(\bth)$\;
  Simulate $\bx^{(i)} \sim p(\bx \mid \bth^{(i)})$\;
  Evaluate distance $\mcal{D}(\bx^{(i)}, \by)$\;
 }
 Retain the $M$ particles $\lbrace{\bth^{(i)}\rbrace}_{i=1}^{M}$ with the lowest losses
 
\caption{Rejection sampling scheme}
\label{alg:Rej}
\end{algorithm}

\subsection{Ricker model}

The Ricker model is a simple model of ecological dynamics that exhibits chaotic behaviour and has an intractable likelihood function. The state of the model, which tracks the size $N_t \in \mathbb{R}_{\geq 0}$ of a population over discrete time steps $t = 1, \dots, n$, evolves as
\begin{equation}
    \log{N_{t+1}} = \log{r} + \log{N_{t}} - N_{t} + \sigma\epsilon_t,
\end{equation}
where $r > 0$ is a growth parameter and $\epsilon_t \sim \mathcal{N}(0, 1)$. Following \citet{Wood2010}, we assume Poissonian observations
\begin{equation}
    \bd{y}_t \sim \text{Po}(\phi N_t) \in \mathbb{N},
\end{equation}
where $\phi > 0$ is a scale parameter. We assume the task of recovering the posterior distribution for $\bth = (\log{r}, \phi, \sigma)$ given a time series of length $n=50$, $\by = (\bd{y}_1, \bd{y}_2, \dots, \bd{y}_{n}) \sim p(\bx \mid \bth^{*})$ with $\bth^{*} = (4, 10, 0.3)$. We take $N_0 = 1$. We further assume the following independent, uniform priors for each parameter:
\begin{equation}
    \log{r} \sim \mathcal{U}(3,8),\quad \quad \quad 
    \phi \sim \mathcal{U}(0,20),\quad \quad \quad 
    \sigma \sim \mathcal{U}(0,0.6).
\end{equation}

\begin{figure}
    \centering
    \includegraphics[width=\linewidth]{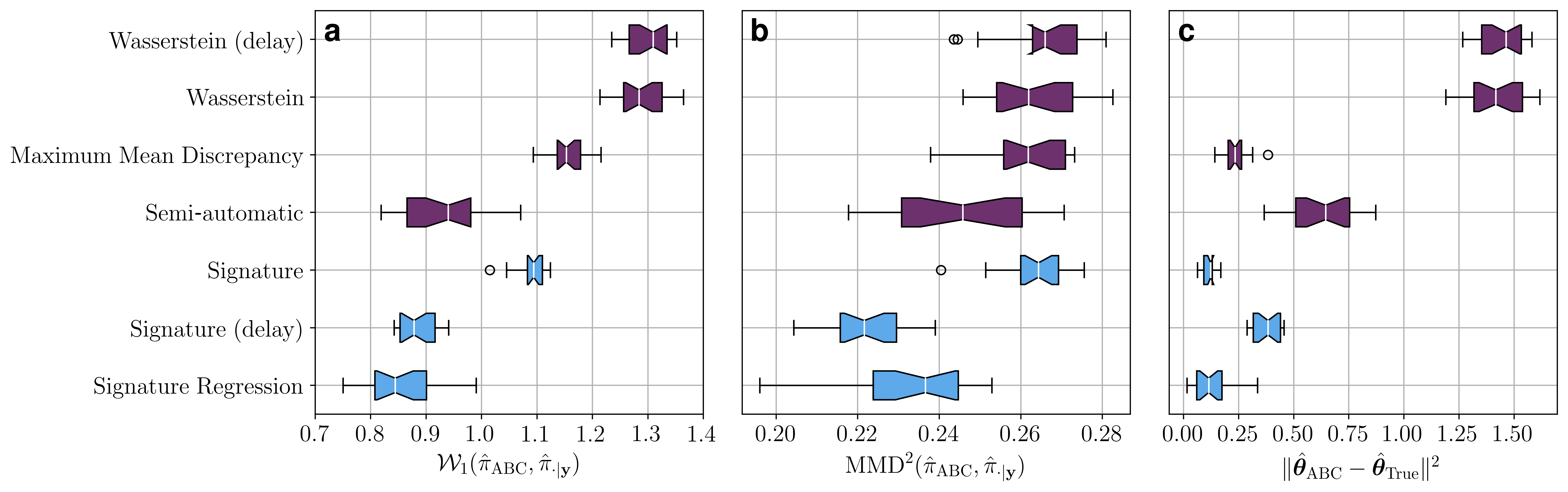}
    \caption{(Ricker model) (\textbf{a}) Wasserstein distances between the posteriors recovered from the different distance measures and an approximate ground truth obtained using \gls{pmcmc}. (\textbf{b}) Maximum mean discrepancies between the posteriors recovered from the different distance measures and an approximate ground truth obtained using \gls{pmcmc}. (\textbf{c}) Squared distances between the means of the \gls{abc} posteriors and the posterior mean obtained using a \gls{pmcmc}. Our methods are shown in blue.}
    \label{fig:ricker_metrics}
\end{figure}

The time series generated by the Ricker model tend to consist of many zero terms, with occasional spikes. For this reason, we use the cumulative sum pre-signature transformation (see Section \ref{sec:PathTrans}) for \gls{sabc}, which is a common transformation for spiking data such as medical data \citep{Morrill2019}. In our experiments, we also found that the \gls{wass}- and \gls{mmd}-based methods benefitted from this transform and were not competitive without it. We therefore also report the results obtained with \gls{wass} and \gls{mmd} with this cumulative sum transform applied. For \gls{saabc}, the hand-crafted summary statistics we use are those proposed in \citet{Wood2010}, and consist of: the autocovariances to lag 5; the mean; the number of zeros in the sequence; the coefficients of the regression $\bd{x}_{t+1}^{0.3} = \beta_1 \bd{x}_{t}^{0.3} + \beta_2 \bd{x}_{t}^{0.6} + \epsilon_t$ for error term $\epsilon_t$; and the coefficients of the cubic regression of the ordered differences $\bd{x}_t - \bd{x}_{t-1}$ on their observed values.

In Figure \ref{fig:ricker_metrics}, we show boxplots for the Wasserstein distances and MMDs between samples from the \gls{abc} posteriors -- denoted with $\hat{\pi}_{\mathrm{ABC}}$ -- and samples from an approximation of the true posterior obtained using \gls{pmcmc} \citep[][see Section \ref{app:mcmc} of the appendix for details]{andrieu2010particle}, which we denote with $\hat{\pi}_{\cdot \mid \by}$. We also show boxplots for the Euclidean distances between the \gls{abc} posterior means and the \gls{pmcmc} posterior mean. These boxplots are all obtained by running the \gls{abc} procedure 20 times with different seeds for each distance measure.

From this, we see that the 
signature-based methods 
tend to produce better performance across all three metrics considered. In more detail, the estimate of the approximate ground truth posterior obtained with the signature-based methods are more accurate than \gls{mmd} and \gls{wass}, as reflected in the Wasserstein distances and MMDs. For \gls{sabc}, this performance gap is enhanced with the additional application of a lag-1 delay transformation (indicated with suffix ``(delay)'' in Figure \ref{fig:ricker_metrics} and subsequent Figures) while no such improvement is observed when applied to \gls{wass}. We note that \gls{saabc} performs particularly well in this example, as a consequence of its use of hand-crafted summary statistics developed specifically for this simulation model. However, the potential power of our signature-based methods is demonstrated by the fact that \gls{skrr} is able to outperform \gls{saabc} in all three metrics, despite the latter using summary statistics carefully engineered by experts. Finally, we 
observe more accurate estimates of the true posterior mean using our signature-based methods than using \gls{wass} and \gls{saabc}, despite the latter using summary statistics carefully engineered by experts to provide accurate inferences for this model. The posterior mean estimates from \gls{sabc} without the delay transformation and \gls{skrr} are also more accurate than those of \gls{mmd}, further evidencing the usefulness of our signature-based methods.

\subsection{Geometric Brownian motion}

\Gls{gbm} is a stochastic differential equation widely used in mathematical finance to model the dynamics of a stock price $x_t$ evolving with time $t$ according to
\begin{equation}
    \mathrm{d} x_t = \mu x_t \mathrm{d}t + \sigma x_t \mathrm{d} W_t,
\end{equation}
where $\mu$ is the percentage drift, $\sigma$ is the volatility, and $W_t$ is a Brownian motion. This model permits an exact discretisation with $i = 1, 2, \dots, n-1$ as
\begin{equation}\label{eq:GBM_disc}
    \log{\bd{x}_{i\Delta t}} = \log{\bd{x}_{(i-1)\Delta t}} + \left(\mu - \frac{1}{2}\sigma^2\right) \Delta t + \sigma \sqrt{\Delta t}\, \epsilon_i,
\end{equation}
which implicitly defines the model $p(\bx \mid \bth)$ from which we simulate. For all simulations, we fix $\bd{x}_0 = 10$, $n = 100$, and $\Delta t = 1/(n-1)$, and simulate the dynamics over the interval $[0,1]$, such that $T=1$.

We consider the task of recovering the posterior for parameters $\bth = (\mu, \sigma)$ given an observation $\by = (\bd{y}_0, \bd{y}_{\Delta t}, \bd{y}_{2\Delta t}, \dots, \bd{y}_{(n-1)\Delta t}) \sim p(\bx \mid \bth^{*} )$ with $\bth^{*} = (0.2, 0.5)$. We assume independent, uniform priors on the parameters as follows:
\begin{equation}
    \mu \sim \mathcal{U}(-1,1),\quad \quad \quad \sigma \sim \mathcal{U}(0.2, 2).
\end{equation}
Inference is amenable to standard, exact likelihood-based Bayesian techniques such as \gls{mh} sampling using the transition density implied by \eqref{eq:GBM_disc}, enabling a comparison against an approximate ground truth posterior. For \gls{saabc}, we follow \citet{Fearnhead2012} and 
regress the parameters $\bth$ onto the first, second, third, and fourth powers of  summary statistics of the time series. Specifically, we take the first, second, third, and fourth powers of the variance and lag-1 and -2 autocorrelations of the increments of the log time series, $\log{(\bd{x}_{i\Delta t}/\bd{x}_{(i-1)\Delta t})}$, since these are informative of the parameters being inferred.

\begin{figure}
\centering
\includegraphics[width=\columnwidth]{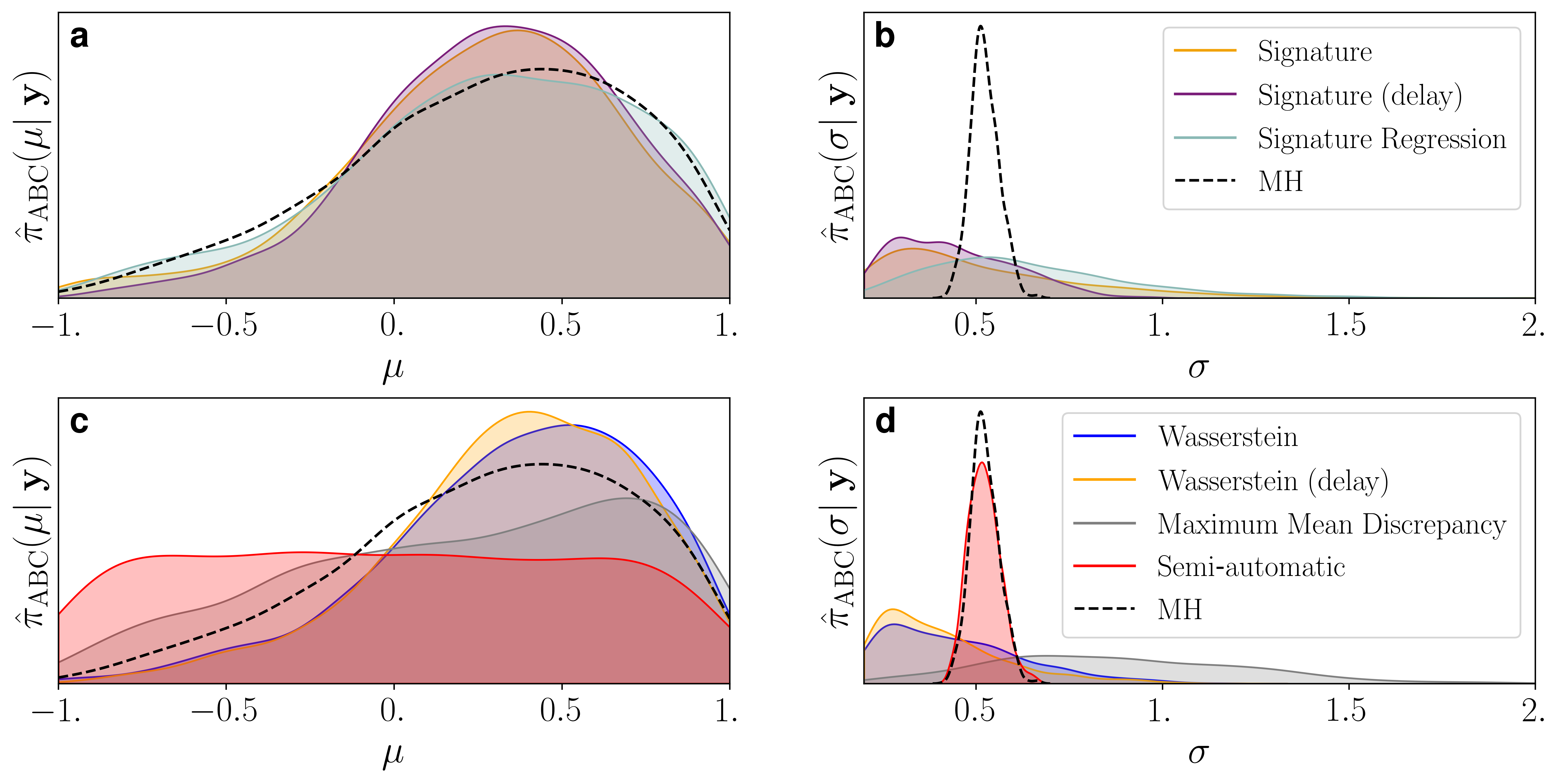}
\caption{(Geometric Brownian motion) Examples of marginal posterior distributions recovered using each loss function and the approximate ground-truth posterior recovered with a Metropolis-Hastings (\gls{mh}) random walk. Panels \textbf{a} and \textbf{b} show the marginal posteriors recovered using our signature methods (\gls{sabc} and \gls{skrr}) and the approximate ground-truth posterior (\gls{mh}). Panels \textbf{c} and \textbf{d} show the marginal posteriors recovered using the Wasserstein distance with curve matching (\gls{wass}), \gls{k2abc} (\gls{mmd}), and semi-automatic \gls{abc} with powers of the variance and lag-1 and -2 autocorrelations of the increments of the log time series as regressors (\gls{saabc}).\label{fig:GBM_marginals}}
\end{figure}

\begin{figure}
    \centering
    \includegraphics[width=\linewidth]{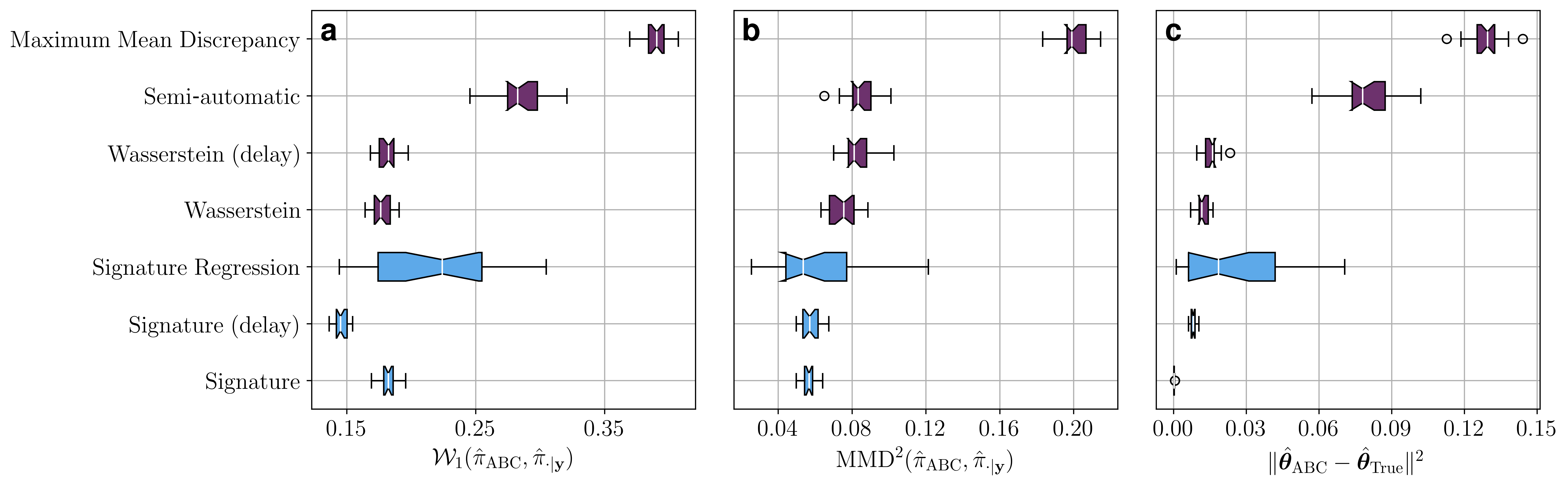}
    \caption{(Geometric Brownian motion) (\textbf{a}) Wasserstein distances between the posteriors recovered from the different distance measures and an approximate ground truth obtained using \gls{mh}. (\textbf{b}) Maximum mean discrepancies between the posteriors recovered from the different distance measures and an approximate ground truth obtained using \gls{mh}. (\textbf{c}) Squared distances between the means of the \gls{abc} posteriors and the posterior mean obtained using \gls{mh}. Our methods are shown in blue.}
    \label{fig:gbm_metrics}
\end{figure}

We show in Figure %
\ref{fig:GBM_marginals} %
the marginal posteriors recovered using the \acrfull{mh} approximation (see Section \ref{app:mcmc} of the appendix for details) and the true likelihood function, along with the approximate posteriors obtained using the rejection sampling scheme in Algorithm \ref{alg:Rej} and each of the distance measures considered. The suffix ``(delay)'' once again indicates that the lag-1 delay transformation was applied. 
From this, we see that and \gls{skrr} and \gls{sabc} track the shape of the approximate ground truth marginal posterior generated by \gls{mh} for $\mu$ more closely than all other methods, and that the marginal distribution for $\sigma$ 
concentrates in the neighbourhood of the approximate ground-truth marginal posterior for $\sigma$. This is in contrast to, for example, the \gls{mmd}, which is overly dispersed and biased for $\sigma$.

In this example, \gls{saabc} has been able to very accurately approximate the marginal density for $\sigma$ as a consequence of the informative set of summary statistics provided to this method. However, \gls{saabc} has experienced difficulty recovering the shape of the marginal density for $\mu$, despite the provided summary statistics also being informative of this parameter. The fact that the signature- and Wasserstein-based methods are able to outperform \gls{saabc}, despite the advantage the latter has been afforded, illustrates the potential power of these methods in cases where the model structure is too complex to easily derive summary statistics that are informative of the parameters.  

In Figure \ref{fig:gbm_metrics}, we show boxplots for the Wasserstein distances and MMDs between the different \gls{abc} posteriors and the approximate ground truth posterior obtained with \gls{mh}, in addition to the Euclidean distance between the \gls{abc} posterior means and the \gls{mh} posterior mean. The boxplots were generated by repeating the \gls{rej} procedure for each distance measure with 20 different random seeds. We see that the superior shape of the signature-based distances also manifests as lower Wasserstein distances and MMDs between their corresponding \gls{abc} posteriors and the \gls{mh} posterior. Indeed, we see that \gls{sabc} with the lag-1 delay transformation uniformly dominates the non-signature methods across all three metrics. 

\subsection{The Brock \& Hommes agent-based model}

In this experiment, we consider a heterogenous agent model proposed by \citet{BROCK19981235} which simulates the dynamics of a set of traders operating under different trading strategies. The system of coupled equations comprising the model may be written succinctly with the following transition density:
\begin{equation*}
    p(\bd{y}_{t+1} \mid \bd{y}_{1:t}, \bth) = \mathcal{N}\left\{f(\bd{y}_{t-2:t}, \bth), \frac{\sigma^2}{R^2}\right\},
\end{equation*}
where
\begin{equation*}
    f\left(\bd{y}_{t-2:t}, \bth\right) = \frac{1}{R}\sum_{j=1}^{J} \frac{\exp{\left\{\beta \left(\bd{y}_t - R \bd{y}_{t-1}\right)\left(g_j \bd{y}_{t-2} + b_j - R \bd{y}_{t-1}\right)\right\}}}{\sum_{j' = 1}^{J} \exp{\left\{\beta \left(\bd{y}_t - R \bd{y}_{t-1}\right)\left(g_{j'} \bd{y}_{t-2} + b_{j'} - R \bd{y}_{t-1}\right)\right\}}}\left(g_j \bd{y}_t + b_j\right)
\end{equation*}
and $R, \beta, \sigma$ are parameters. In this way, we are able to obtain an approximate ground truth posterior with standard \gls{mcmc} techniques such as \gls{mh}. We follow \citet{PLATT2020103859, dyer2022black} and assume the following parameter values: $J=4, R = 1.0, \sigma = 0.04, \beta = 10, g_1 = b_1 = b_4 = 0$ and $g_4 = 1.01$. 

The parameters $g_j \in \mathbb{R}$ capture the trend-following tendencies of the agents, while the parameters $b_j \in \mathbb{R}$ determine the biases towards different trading strategies. In our experiments, we consider the task of estimating the posterior $\pi\left(\bth \mid \by\right)$, where $\bth = \left(g_2, b_2, g_3, b_3\right)$, $\by = (\bd{y}_1, \dots, \bd{y}_n) \sim p(\bx \mid \bth^{*})$ is the pseudo-observation, $n=100$, and $\bth^{*} := (-0.7, -0.4, 0.5, 0.3)$ is the parameter setting used to generate $\by$. 

\begin{figure}
    \centering
    \includegraphics[width=\linewidth]{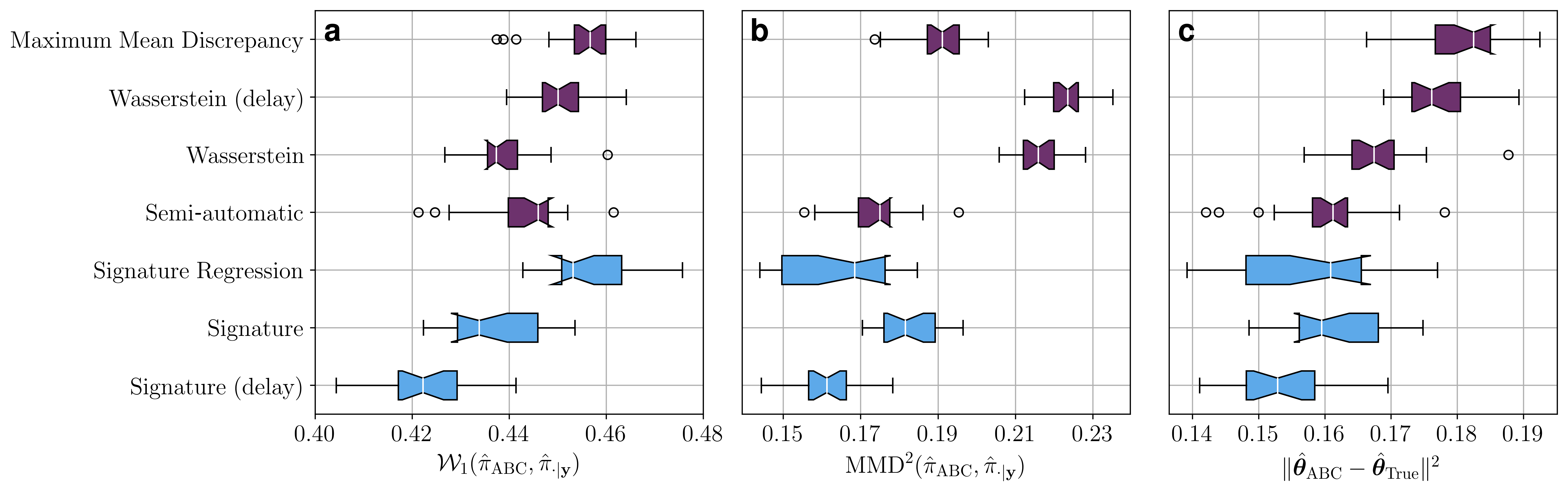}
    \caption{(Brock \& Hommes) (\textbf{a}) Wasserstein distances between the posteriors recovered from the different distance measures and samples from the exact posterior. (\textbf{b}) Maximum mean discrepancies between the posteriors recovered from the different distance measures and samples from the exact posterior. (\textbf{c}) Squared distances between the means of the \gls{abc} posteriors and the exact posterior mean. Our methods are shown in blue.}
    \label{fig:bhn_metrics}
\end{figure}

We show in Figure \ref{fig:bhn_metrics} boxplots for the Wasserstein distance and MMD between the \gls{abc} posteriors, denoted with $\hat{\pi}_{\mathrm{ABC}}$, and the approximate ground-truth posterior obtained with \gls{mh}, denoted with $\hat{\pi}_{\cdot\mid\by}$. We also show boxplots for the Euclidean distance between the \gls{abc} posterior means and the \gls{mh} posterior mean. These boxplots were created by running the \gls{rej} algorithm with the same 20 random seeds. In this experiment, \gls{saabc} uses the first and second powers of $l$ evenly spaced order statistics of the output data $\bx$, as considered in \citet{Fearnhead2012}, where we take $l=10$.

From this, we see that the signature-based methods tend to generate lower values in all three metrics compared to existing methods. In particular, we see that \gls{sabc} with the lag-1 delay transformation once again dominates existing methods uniformly across all three metrics, while the same transformation applied to \gls{wass} does not result in the same improvement. This demonstrates the potential power of our signature-based methods as automatic distance measures for \gls{abc} for dynamic, stochastic simulators.

\subsection{An example of irregular, multivariate data: generalised stochastic epidemics}

As previously discussed, the signature method naturally allows for inference with multivariate and/or irregularly spaced time series. To demonstrate this, we consider a generalised stochastic epidemic model \citep{Kypraios2007EfficientBI}, which simulates the spread of an infection through a fixed population of $N$ individuals. Individuals are initially susceptible, may become infected, and subsequently recover without the possibility of reinfection. Dynamics of the model are determined by parameters $\beta$ and $\gamma$, which control the rate of infection and recovery according to the following transition probabilities:
\begin{align}
    P\left(X_{t + \delta t} - X_t = -1, Y_{t+\delta t} - Y_t = 1 \mid \mathcal{H}_t \right) &= \beta X_t Y_t \delta t + o(\delta t),\\
    P\left(X_{t + \delta t} - X_t = 0, Y_{t+\delta t} - Y_t = -1 \mid \mathcal{H}_t \right) &= \gamma Y_t \delta t + o(\delta t),\\
    P\left(X_{t + \delta t} - X_t = 0, Y_{t+\delta t} - Y_t = 0 \mid \mathcal{H}_t \right) &= 1 - \beta\, X_t Y_t \delta t + \gamma Y_t \delta t + o(\delta t),
\end{align}
where $X_t$ and $Y_t$ are the number of susceptible and infected individuals at time $t \in [0, T]$, respectively, and $\mathcal{H}_t$ is a sigma-algebra generated by the process up until time $t$. These three transition probabilities thus capture infection, recovery, and an absence of activity, respectively. 

We consider the problem of recovering the posterior density for $\bth = (\beta, \gamma)$ given observations of the infections and recoveries occurring in the observation period $[0, T]$ with $T=50$ in a system of $Z=100$ individuals. For every simulation, the epidemic begins with one infected individual at time $t=0$. We generate ``empirical'' data at parameters $\bth^{*} = (10^{-2}, 10^{-1})$, and assume Gamma priors for both $\beta$ and $\gamma$,
\begin{equation}
    \beta \sim \Gamma(\lambda_{\beta}, \nu_{\beta}),\quad \quad \quad \gamma \sim \Gamma(\lambda_{\gamma}, \nu_{\gamma}),
\end{equation}
with $\lambda_{\beta} = 0.1$, $\nu_{\beta} = 2$, $\lambda_{\gamma} = 0.2$, and $\nu_{\gamma} = 0.5$. It can be shown \citep{Kypraios2007EfficientBI} that this prior is conjugate for the model, leading to the posterior density
\begin{multline}\label{eq:GSE_post}
    \pi(\beta, \gamma \mid \mathbf{I}, \mathbf{R}) \propto \beta^{\lambda_{\beta} + n_{I} - 2} \exp\left\{ -\beta \left(\int_{\phi_1}^{T} X_t Y_t\, {\mathrm d} t + \nu_{\beta}\right) \right\}\\
    \gamma^{\lambda_{\gamma} + n_{R} - 1 }\exp\left\{ -\gamma \left(\int_{\phi_1}^{T} Y_t\, {\mathrm d} t + \nu_{\gamma}\right) \right\},
\end{multline}

where $\mathbf{I}$ and $\mathbf{R}$ are the infection and recovery times, respectively, $n_I$ and $n_R$ are the total number of individuals in the model that are infected and that recover over the course of the simulation, respectively, and $\phi_1$ is the time of the first infection. Thus, samples can be drawn from the exact posterior for a given dataset simulated by this model. 
We simulate the model using the Gillespie algorithm \citep{Gillespie}, such that the lengths of the simulated sequences, and the spacing between points in the sequences, are both also random. 

To perform \gls{sabc}, we bring all three channels of the multivariate stream --- the number of infected individuals, number of recovered individuals, and time --- into the range $[0,1]$ by dividing by $Z$, $Z$, and $T$, respectively. For \gls{wass}, we set $\lambda = 2$, since the expected vertical range is approximately twice that of the horizontal range $T=50$ when $Z=100$.

\begin{figure}
    \centering
    \includegraphics[width=\linewidth]{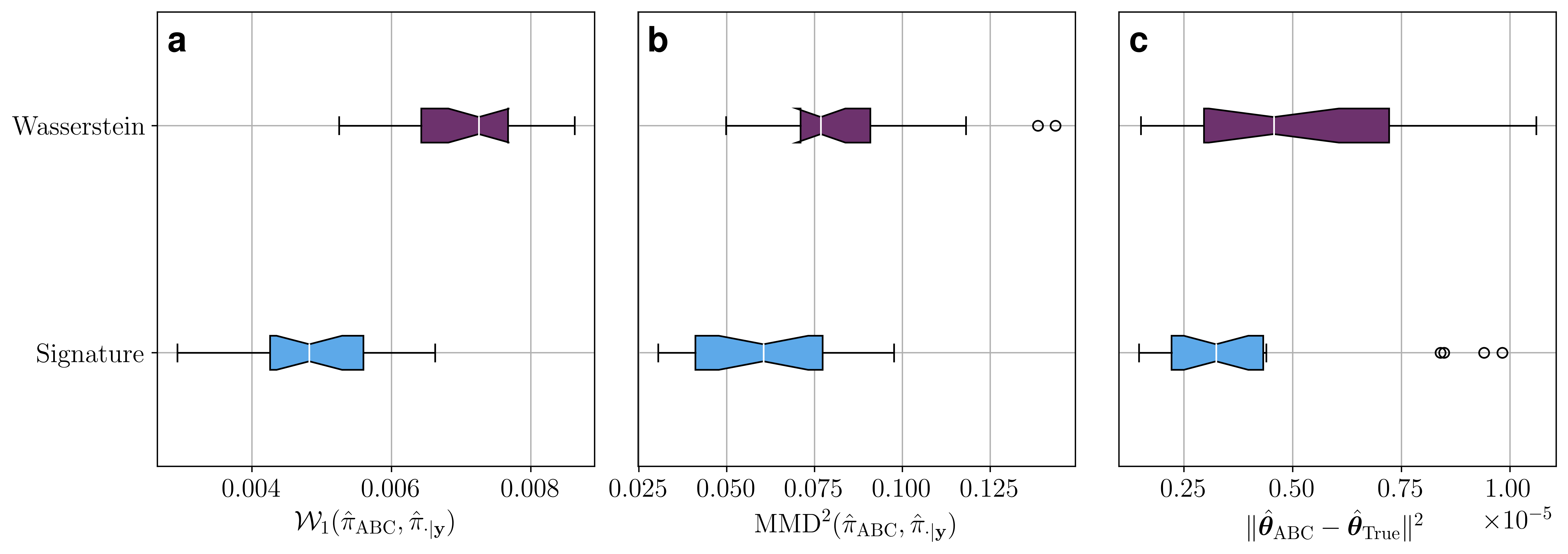}
    \caption{(Generalised stochastic epidemic model) (\textbf{a}) Wasserstein distances between the posteriors recovered from the different distance measures and samples from the exact posterior. (\textbf{b}) Maximum mean discrepancies between the posteriors recovered from the different distance measures and samples from the exact posterior. (\textbf{c}) Squared distances between the means of the \gls{abc} posteriors and the exact posterior mean. Our method is shown in blue.}
    \label{fig:gse_metrics}
\end{figure}

\begin{figure}
\centering
\includegraphics[width=0.7\columnwidth]{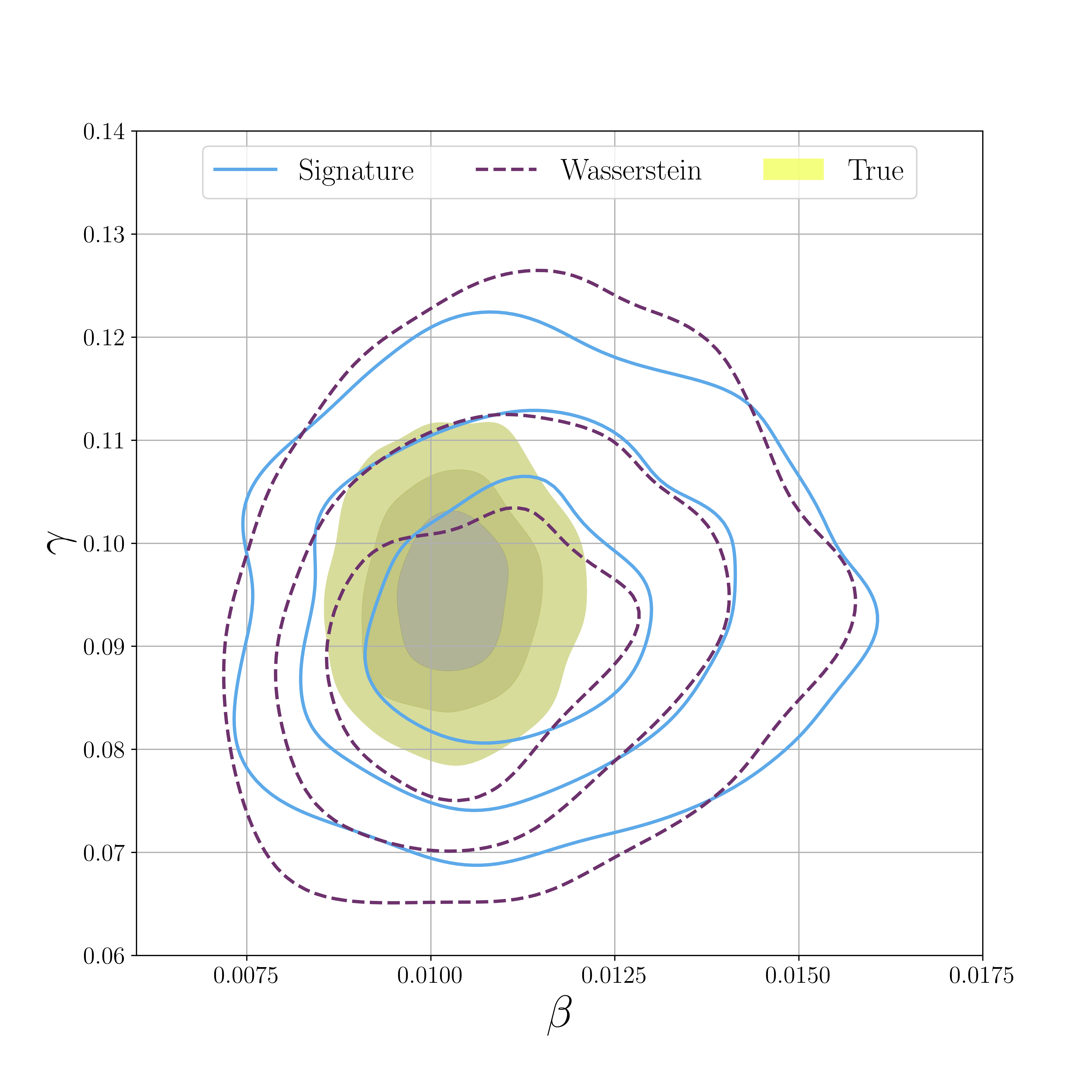}
\caption{(Generalised stochastic epidemic model) A contour plot of the joint posterior densities recovered with the Wasserstein distance (dashed purple lines) and Signature \gls{abc} (solid blue lines), and samples from the exact posterior (filled yellow contours).\label{fig:GSE_post}}
\end{figure}

We show in Figure \ref{fig:gse_metrics} boxplots for the Wasserstein distances and MMDs between samples from \gls{wass} and \gls{sabc} posteriors and samples from the exact posterior. To obtain these approximate posteriors, we run Algorithm \ref{alg:Rej} with $N=10^5$ and $M=100$ for 20 different seeds. We also show boxplots for the distribution of squared distances between the posterior means obtained with \gls{wass} and \gls{sabc} and the exact posterior mean. (In this experiment, we observed the \gls{abc} posterior obtained with the \gls{mmd} distance measure to perform considerably worse than \gls{wass} and \gls{sabc}, and therefore omit these results from Figure \ref{fig:gse_metrics} for clarity.) We also show contour plots obtained by running the inference procedure at these 20 different seeds and pooling the best $M$ losses from each in Figure \ref{fig:GSE_post}, along with samples from the exact posterior, \eqref{eq:GSE_post}. 

From all of this, we see that the natural notion of distance between multivariate and irregularly sampled time series data of different lengths, enabled by the use of path signatures, manifests as better recovery of both the true posterior distribution and the true posterior mean in this example, in which the Wasserstein distances and MMDs between posteriors and Euclidean distances between posterior means for \gls{sabc} are generally lower than those obtained using \gls{wass}. 

\begin{figure}
    \centering
        \hfill
        \includegraphics[width=\columnwidth]{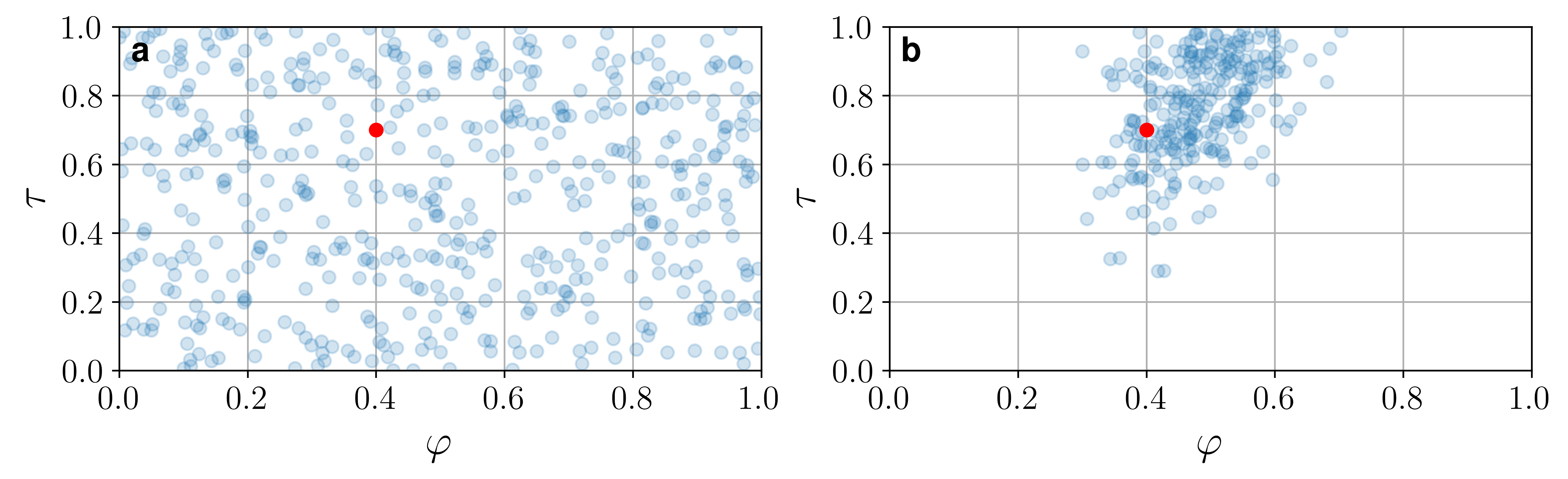}
    \caption{(Dynamic graph model) Samples from (\textbf{a}) the prior and (\textbf{b}) the posterior obtained from \gls{sabc} for the dynamic graph model.}\label{fig:graphs}
\end{figure}

\subsection{A dynamic graph model}

In the previous experiments, we have seen that our signature-based methods are able to outperform existing approaches to \gls{abc} for time series simulators that generate both regularly spaced, univariate sequences and irregularly spaced, multivariate sequences of random length. However, a further consequence and benefit of the kernelisation of our signature-based approaches is that such methods can be applied to more exotic problems, in which the data evolves in more general topological spaces. For example, equipped with a suitable kernel on graphs, we may apply our signature-based methods to parameter inference problems that arise for \textit{dynamic graph simulators} that have intractable likelihood functions.

As an illustration of this point, we take as a final example a simple dynamic graph model described in \citet{zhang2017random}, which can be seen as the dynamic counterpart to the canonical Erd\H{o}s-R\'{e}nyi random graph model \citep{erdos1959, erdos1960}. In this model, edges appear with probability $\varphi$ at time $t=1, \dots, n$ where they were absent at time $t-1$, or remain absent with probability $1 - \varphi$. Similarly, edges that were present at time $t-1$ disappear with probability $\tau$ at time $t$ or remain present with probability $1-\tau$. The output of the simulator can thus be taken as, for example, the sequence of graph snapshots or, equivalently, their adjacency matrices $\bd{A}_{t}$ in which $[\bd{A}_t]_{ij} =$ the number of times edge $(i,j)$ has appeared across all time steps $t'=0,\dots,t$, where $\bd{A}_0$ is some initial seed network.

We consider the task of estimating the posterior $\pi(\bth \mid \bd{A})$ for parameters $\bth := (\varphi, \tau)$ given some observation $\bd{A} := (\bd{A}_0, \bd{A}_1, \dots, \bd{A}_n) \sim p(\bd{B}\mid \bth^{*})$, where $n=25$, $\bd{A}_t \in \mathbb{R}^{20\times 20}$, and $\bth^{*} = (0.4,0.7)$ are the generating parameters. We assume uniform priors $\varphi \sim \mathcal{U}(0,1)$, $\tau \sim \mathcal{U}(0,1)$. 
We time-augment by using the product\footnote{Such tensor product kernels are valid kernels on product spaces.} of a \gls{wl} kernel \citep{shervashidze2011weisfeiler} on graphs $\bd{A}_t$ and a Gaussian RBF kernel on the time-channel for $\kappa$:
\begin{equation}\label{eq:graph_kernel}
    \kappa\left\{\left(\bd{A}_t, t\right), \left(\bd{B}_s, s\right)\right\} = \text{WL}(\bd{A}_t, \bd{B}_s) \cdot \exp^{-\frac{\| t - s \|^2}{\sigma}},
\end{equation}

in which 
the initial labels for all nodes in each graph in all sequences is taken to be identically $1$. Furthermore, we perform two iterations of the message-passing and hashing procedure, and use a vertex histogram kernel as the base kernel. 

We show the posterior we obtain from Algorithm \ref{alg:Rej} -- using $N=10^5$, $M=250$, and the signature distance \eqref{eq:sig_distance} using \eqref{eq:graph_kernel} as the static kernel $\kappa$ -- in Figure \ref{fig:graphs}. From this we see that 
the \gls{sabc} posterior has been able to concentrate significantly around the generating parameters $\bth^{*}$, 
suggesting that our signature-based approach can furthermore be successfully applied to simulators generating data evolving in more general topological spaces than $\mathbb{R}^d$. 

\section{Conclusion}

In this paper, we introduced two novel approaches---Signature \gls{abc} and Signature Regression \gls{abc}---to performing approximate Bayesian computation with time series simulation models. Each method relies on the path signature---an object that is fundamental to the theory of controlled differential equations and rough paths---and that is associated with the path traversed by a sequence of data points. In particular, we make use of the recently developed signature kernel to construct and compute discrepancies between time series data arising in \gls{abc} settings without manually contriving summary statistics. 

We show that the natural notion of distance between time series to which such an approach leads satisfies conditions under which the \gls{abc} posterior converges to the ground-truth posterior (under certain regularity conditions on the simulator's likelihood function) and discuss the robustness properties of the Signature \gls{abc} posterior as the number of data points $n\to \infty$ within a finite time horizon for a fixed \gls{abc} tolerance parameter. As an illustration of our proposed methods, we present multiple examples of Bayesian inference tasks in which our approaches outperform existing techniques that are common in the approximate Bayesian inference literature; indeed, in each experiment we consider, at least one signature-based method uniformly dominates competing methods across all three of the metrics considered in this paper. We demonstrate that our methods flexibly accommodate a number of potentially helpful transformations of the data---for example, delay transformations---and in our final examples that our methods are applicable to more complex settings than univariate time series, for example multivariate and irregularly sampled sequences and even simulators that generate non-Euclidean time series.

While we have compared the different distance measures using a basic rejection algorithm in this paper in order to allow for a simple and transparent comparison, we note that our proposed methods can be embedded within other more sophisticated sampling algorithms, for example \gls{mcmc} or sequential Monte Carlo methods. Additionally for the Signature Regression \gls{abc} method, there is the possibility of incorporating mechanisms for generating more accurate regression results, for example using a pilot run to determine regions of non-negligible posterior mass as described in \citet{Fearnhead2012}. This may allow for improved approximations to the true posterior density.

\subsection{Future work}

Throughout the above, we have assumed that only one sequence $\bd{y}$ has been observed from the real world. This is a realistic assumption in many useful real-world cases; for example, this is often the case in macroeconomics or during a pandemic, where it would be incorrect to treat signals recorded at e.g. the country level as being \textit{iid} rather than as different channels in a single observed sequence. 

However, there are certain realistic settings in which multiple sequences $\lbrace{\bd{y}^{(j)}}\rbrace_{j=1}^{J}$ are recorded in which an \textit{iid} assumption is reasonable. For example, in healthcare settings, recordings of patients with similar medical profiles may reasonably be modelled as \textit{iid} draws from some underlying distribution. Similarly, in the natural or behavioural sciences, it is sometimes possible to perform multiple trials or repetitions of experiments in which the evolution of some quantity is recorded. In these cases, the following two generalisations of the approach taken in this chapter may be useful:
\begin{enumerate}
    \item[(a)] taking $\mcal{P}^J = \lbrace{\bd{y}^{(j)}}\rbrace_{j=1}^{J}$, we may use the discrepancy measure
    \begin{align}\nonumber
    \Ds(\delta_{\bx}, \mcal{P}^J) &:= \norm{\mathbb{E}_{\bx \sim \delta_{\bx}}(\sig{\bx}) - \mathbb{E}_{\by \sim \mcal{P}}(\sig{\by})}{}^2 \\
    &= k(\bx, \bx) + \frac{1}{J(J-1)} \sum_{i \neq j} k(\by^{(i)}, \by^{(j)}) - \frac{2}{J}\sum_{j = 1}^J k(\bx, \by^{(j)})
    \end{align}
    each time we query a new parameter $\bth$ -- where $\delta_{\bd{x}}$ is a point mass located on $\bd{x} \sim p(\cdot \mid \bth)$ -- as the distance measure in \gls{abc}. This provides a meaningful comparison between a single output from the dynamic, stochastic simulator and the empirical measure on sequences given by the real-world dataset when the simulation budget should be kept as low as possible;
    \item[(b)] more generally, when there is greater tolerance for a larger simulation burden, one may instead simulate $N \geq 1$ times at each $\bth$ to construct an empirical measure $\mcal{P}^N_{\bth} = \lbrace{\bd{x}^{(n)}}\rbrace_{n=1}^{N}, \bd{x}^{(n)} \overset{iid}{\sim} p(\cdot \mid \bth)$ and use the full \gls{mmd} between (in general non-Dirac) measures on sequences:
    \begin{align}\nonumber
        \Dm(\mcal{P}^N_{\bth}, \mcal{P}^J) &:= \norm{\mathbb{E}_{\bx \sim \mcal{P}^N_{\bth}}(\sig{\bx}) - \mathbb{E}_{\by \sim \mcal{P}^J}(\sig{\by})}{}^2\\\nonumber
        &= \frac{1}{N(N-1)} \sum_{i \neq j} k(\bx^{(i)}, \bx^{(j)}) + \frac{1}{J(J-1)} \sum_{i \neq j} k(\by^{(i)}, \by^{(j)})\\
        &\quad \quad \quad \quad \quad \quad - \frac{2}{NJ} \sum_{i,j} k(\bx^{(i)}, \by^{(j)}).
    \end{align}
\end{enumerate}

The latter of these may also be useful in the case of a single observation and simulation with $J=N=1$ in the following way: if the data-generating process is known to be ergodic, it may be reasonable to treat successive blocks/sub-sequences of $\by$ and $\bx$ as being approximately \textit{iid}. Then, taking $\mcal{P}^N_{\bth}$ and $\mcal{P}^J$ to be the empirical measures associated with the collection of blocks of $\bx$ and $\by$, respectively, $\Dm$ provides a reasonable discrepancy to be used in \gls{abc} for dynamic, stochastic simulation models with intractable likelihood functions.

\section{Data availability statement}

All data and code for reproducing the experimental results presented in this manuscript are available in the supplementary material and on GitHub at \url{https://github.com/joelnmdyer/SignatureABC}, and can be accessed with \url{https://doi.org/10.5281/zenodo.7246198}.

\section{Acknowledgements}

The authors are grateful to Horatio Boedihardjo, Lajos Gergely Gyurko, Zacharia Issa, Terry Lyons, James Morrill, Harald Oberhauser, and Cristopher Salvi for their comments, feedback, and helpful discussions. JD was supported by the EPSRC Centre For Doctoral Training in Industrially Focused Mathematical Modelling (EP/L015803/1) in collaboration with Improbable. JD was also supported by The Alan Turing Institute under the EPSRC grant EP/N510129/1.

\bibliography{references}

\appendix

\appendix

\section{Path signatures}\label{app:signatures}

\subsection{Further background on path signatures}

To introduce signatures more completely, it is instructive to consider a simple example of a finite-dimensional path:

\begin{Example}[Example 2.3, \citet{Kiraly2019}]
    Let $h_t$ take values in $\mathbb{R}^2$, $h_t = (a_t, b_t)$. Then
    \begin{equation*}
        S_1(h) = 
        \left[\begin{array}{CC}
            \int_0^T {\d }a_t\\[4pt] \int_0^T {\d }b_t 
        \end{array}\right]
        \ \ \ 
        \text{and}\ \ \ 
        S_2(h) =         
        \left[\begin{array}{CC}
            \int_0^T \int_0^{t_2} {\d }a_{t_1}{\d }a_{t_2} & \int_0^T \int_0^{t_2} {\d }a_{t_1}{\d }b_{t_2} \\[4pt]
            \int_0^T \int_0^{t_2} {\d }b_{t_1}{\d }a_{t_2} & \int_0^T \int_0^{t_2} {\d }b_{t_1}{\d }b_{t_2}
        \end{array}\right].
    \end{equation*}
    These terms can be further interpreted geometrically: the terms in $S_1(h)$ capture the increments along each dimension, while the off-diagonal elements of $S_2(h)$ capture the areas above and below the curve; see Figure \ref{fig:geom}. Higher order terms capture higher order notions of area that are more difficult to visualise and interpret.

    \begin{figure}[ht]
        \centering
        \includegraphics[width=0.75\linewidth]{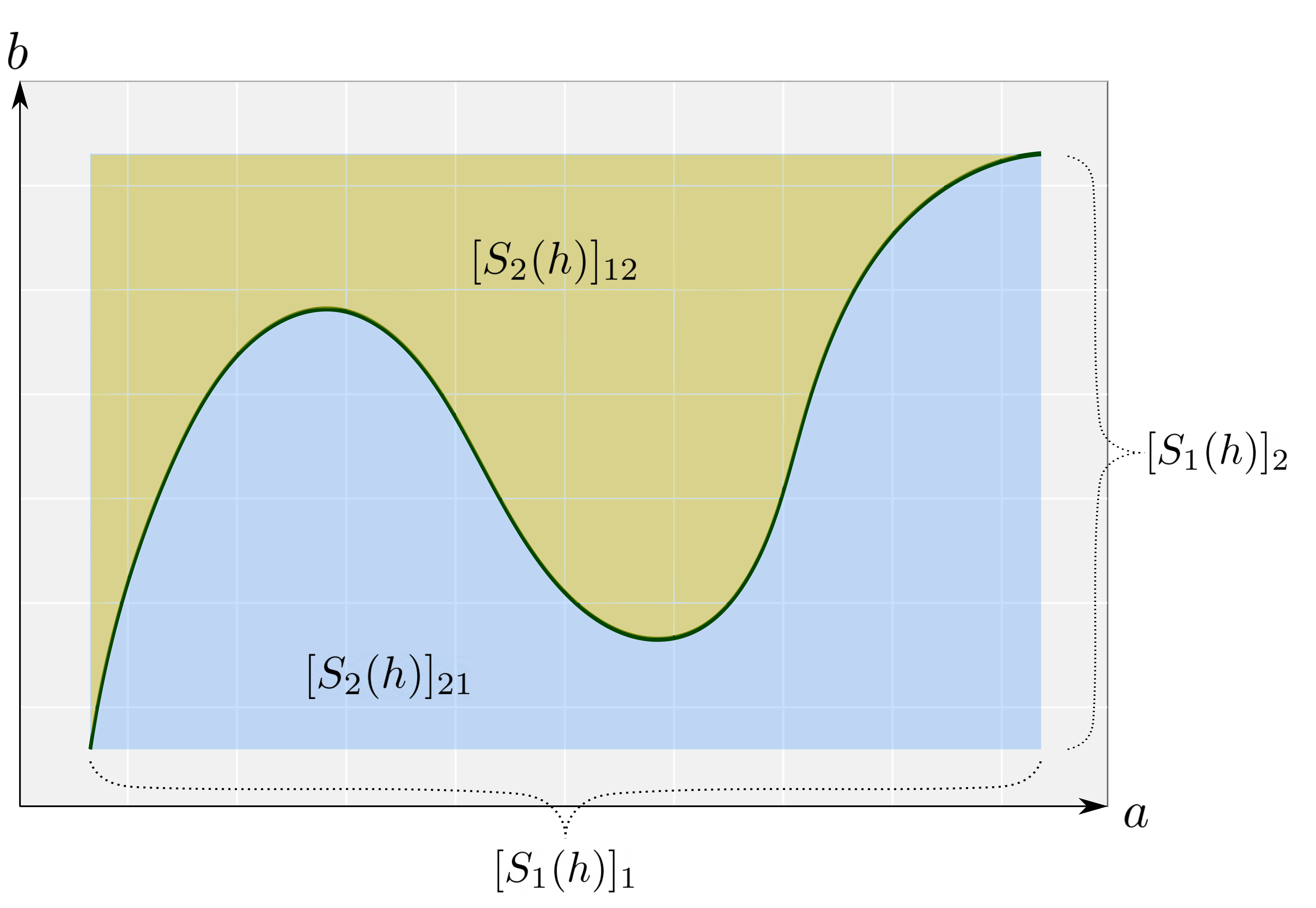}
        \caption{Geometric interpretation of the signature terms for an example two-dimensional path, shown as the dark green curve. Depth-1 terms correspond to the increments $a_T - a_0$ and $b_T - b_0$, while the depth-2 terms $[S_2(h)]_{21}$ and $[S_2(h)]_{12}$ correspond to the blue and yellow areas, respectively.}\label{fig:geom}
    \end{figure}
    
\end{Example}

\begin{Remark}
    Since we have assumed our paths to be of bounded variation, the integrals above can be understood as the Riemann-Stieljes integrals with respect to $h$. When the underlying path is not smooth, the integrals are taken to be stochastic or rough path integrals \citep{chevyrev2018signature}. For example, in the case of Brownian motion in $\mathbb{R}^d$, the integrals are stochastic and can be taken in the Stratonovich sense. For a larger class of stochastic processes, rough path theory \citep{LyonsT.J2007Dedb} provides an integration theory that enables the computation of the terms in the signature. As we will discuss later, this work considers throughout only linear interpolations between points in time series, so all paths considered here are of finite variation.
\end{Remark}

Path signatures are thus infinite sequences of statistics for path-valued random variables capturing information regarding the order of observations along, and the interaction between, different channels of the path. They are grounded in the theory of \glspl{cde} and stochastic analysis, and appear in the solutions of \glspl{cde} and \glspl{sde} as obtained through a procedure analogous to Picard iterations for ordinary differential equations.

To see this, we follow \citet{LyonsT.J2007Dedb} and let $V$ and $W$ be two Banach spaces, $B : V \to \mathbf{L}(W, W)$ be a bounded linear map -- where $\mathbf{L}(W, W)$ denotes the space of bounded linear mappings from $W \to W$ --- and $h : [0,T] \to V$ be a continuous path of bounded variation. Consider the following set of linear equations:
\begin{flalign}\label{eq:cde_1}
    && {\d }g_t &= Bg_t\, {\d }h_t,\ \ g_0 \in W &&\\\label{eq:cde_2}
    && {\d }\phi_t &= B\phi_t\, {\d }h_t,\ \ \phi_0 \in \mathbf{L}(W, W). &&
\end{flalign}
Here, $Bg_t\, {\d }h_t$ is taken to mean $\left\{B({\d }h_t)\right\}(g_t)$ while $B\phi_t\, {\d }h_t$ is $B({\d }h_t)\circ \phi_t$. By applying the aforementioned iterative procedure to recover the solution $\phi_t$ to \eqref{eq:cde_2}, we obtain
\begin{equation}\label{eq:cde_sol}
    \phi_t = \sum_{m \geq 0} B^{\otimes m} \int_{0}^t {\d }h^{\otimes m},
\end{equation}
in which we see that the signature terms, Equation \eqref{eq:sig_terms}, appear in the summand. The solution to \eqref{eq:cde_1} is then obtained from the flow $\phi_t$ as $g_t = \phi_t(h_0)$. Similarly, a solution to the following linear \gls{sde} driven by Brownian motion $W$,
\begin{equation*}
    \d Y_t = A(Y_t) \circ \d W_t, \quad Y_0 = y_0
\end{equation*}
for some linear operator $A$, can be obtained as
\begin{equation*}
    Y_t = \sum_{m \geq 0} A^{\otimes m} S_{m,[0,t]}(W)\, y_0,
\end{equation*}
where $S_{m,[0,t]}(W)$ is the order-$m$ tensor in the signature of $W_t$ over interval $[0,t]$ and the integrals are taken in the Stratonovich sense \citep[][Section 3.3.2]{LyonsT.J2007Dedb}. As we have seen here, signatures arise naturally as good approximations to solutions of \glspl{cde} and \glspl{sde}, and accurately describe the response of systems such as that of Equations \eqref{eq:cde_1}-\eqref{eq:cde_2} to an input signal $h$, where the inclusion of terms of increasing order further refine the approximate solution. The above sums, such as in Equation \eqref{eq:cde_sol}, converges as a result of the factorial rate of decay of the terms in the signature:
\begin{Proposition}[Proposition 2.2, \citet{LyonsT.J2007Dedb}]\label{thm:fact_decay}
    Let $V$ be a Banach space and $h \in \bv{V}$. Then, for each $m \geq 0$,
    \begin{equation}
        \norm{\int_{0}^{T} {\d }h^{\otimes m}}{V^{\otimes m}} \leq \frac{\onevar{h}}{m!}.
    \end{equation}
\end{Proposition}

\begin{Remark}
The signature of a univariate path consists only of powers of the difference between the final and initial points in the stream \citep[see e.g.][Example 5]{Chevyrev2016}. Therefore in practice one always considers paths in at least two dimensions. This can always be achieved by including the observation time as a channel in the path.
\end{Remark}

\subsection{Shuffle-product property}\label{app:shuffle_sig}

The terms of the path signature exhibit a so-called \textit{shuffle-product} property: 

\begin{Theorem}[Theorem 2.29, \citet{LyonsT.J2007Dedb}]
    Let $h \in \bv{\mcal{H}}$. Then
    \begin{equation*}
        \int_{0}^T {\d }h^{\otimes m} \otimes \int_{0}^T {\d }h^{\otimes m'} = \sum_{\sigma} \sigma\left(\int_{0}^T {\d }h^{\otimes (m+m')} \right),
    \end{equation*}
    where the sum is taken over all order shuffles, defined as
    \begin{multline*}
        \{\sigma: \sigma \text{ is a permutation of } \lbrace{1, \dots, m + m'\rbrace}\\ \text{ with } \sigma(1) < \dots < \sigma(m), \sigma(m+1) < \dots < \sigma(m+m')\}.
    \end{multline*}
    $\sigma$ then acts on $\mcal{H}^{\otimes (m+m')}$ as $\sigma(e_{i_1} \otimes \dots \otimes e_{i_{m+m'}}) = e_{\sigma(i_1)} \otimes \dots \otimes e_{\sigma(i_{m+m'})}$.
\end{Theorem}

\subsection{Additional pre-processing}

One further and sometimes desirable pre-processing step is the \textit{lead-lag transformation}:

\paragraph{Lead-lag transformation} This transformation operates on a sequence $\bd{x} = (\bd{x}_{t_1}, \bd{x}_{t_2}, \dots, \bd{x}_{t_n})$ as follows:
\begin{equation}
    (\bd{x}_{t_1}, \bd{x}_{t_2}, \dots, \bd{x}_{t_n}) \mapsto \left((\bd{x}_{t_1}, \bd{x}_{t_1}), (\bd{x}_{t_1}, \bd{x}_{t_2}), (\bd{x}_{t_2}, \bd{x}_{t_2}), \dots, (\bd{x}_{t_{n-1}}, \bd{x}_{t_n}), (\bd{x}_{t_n}, \bd{x}_{t_n})\right).
\end{equation}
Under this transformation, the number of channels in the sequence doubles, and the sequence length increases from $n$ to $2n-1$. Applying this transformation enables the signature to emphasise certain properties of the path %
such as the quadratic variation %
and the L\'{e}vy area when combined with the cumulative sum \citep{Gyurk2014, Chevyrev2016}. For datasets for which these quantities are believed to be important, applying the lead-lag transformation may be appropriate.

\section{Background on rough path theory}\label{app:rough_paths}

In this section, we provide some basic definitions and results in the theory of rough paths that are used or discussed in the main text. Throughout this section, $V$ will be a Banach space and $\Delta_{[0,T]} := \lbrace{(s,t) \in [0,T]^2 : 0\leq s \leq t\leq T}\rbrace$.

\begin{Definition}[Tensor algebra]
    The \emph{truncated tensor algebra} at integer degree $n$ over $V$
    \begin{equation*}
        T^{(n)}(V) := \left\{ s \in \prod_{k = 0}^{n} V^{\otimes k} := \mathbb{R} \oplus V \oplus (V \otimes V) \oplus \dots \oplus V^{\otimes n}\ \Bigg\vert\ s_0 = 1 \right\},
    \end{equation*}
    where $s_0$ indicates the first element of $s \in T^{(n)}(V)$. The \emph{extended tensor algebra} $T((V))$ is the infinite sequence $\prod_{n \geq 0}V^{\otimes n}$.
\end{Definition}

With this definition in place, we can now define a multiplicative functional.

\begin{Definition}[Multiplicative functional, Definition 3.1 of \citet{LyonsT.J2007Dedb}]
    Let $n\geq 1$ be an integer, and $X : \Delta_{[0,T]} \to T^{(n)}(V)$ be a continuous map. For each $(s,t)\in \Delta_{[0,T]}$, denote by
    \begin{equation*}
        X_{s,t} := (X^0_{s,t}, X^1_{s,t}, \dots, X^n_{s,t}) \in \prod_{k = 0}^{n} V^{\otimes k}
    \end{equation*}
    the image of $(s,t)$ under $X$. If $X^0_{s,t} = 1\ \forall (s,t) \in \Delta_{[0,T]}$ and
    \begin{equation*}
        X_{s,u} = X_{s,t} \otimes X_{t,u}\quad \forall s,t,u \in [0,T]\ \text{s.t. } s \leq t \leq u,
    \end{equation*}
    then $X$ is called a \emph{multiplicative functional} of degree $n$ in $V$.
\end{Definition}

\begin{Remark}
    The path signature for a bounded variation path $X : [0,T] \to V$ truncated to some finite degree $M$ is an element of the truncated tensor algebra at degree $M$ over $V$, and is a multiplicative functional as a result of Chen's identity \citep{chen1958integration}, giving that
    \begin{equation*}
        S^{\leq M}_{s,u} = S^{\leq M}_{s,t} \otimes S^{\leq M}_{t, u}\quad \forall s,t,u \in [0,T]\ \text{s.t. } s\leq t\leq u,
    \end{equation*}
    where $S^{\leq M}_{s,t}$ denotes the collection of the first $M$ tensors in the signature integrated over $(s,t) \in \Delta_{[0,T]}$.
\end{Remark}

\begin{Definition}[Rough path, Definition 3.11 of \citet{LyonsT.J2007Dedb}]
    Let $p \geq 1$ be a real number. A $p$-rough path in $V$ is a multiplicative functional of degree $\lfloor{p}\rfloor$ in $V$ with finite $p$-variation. The space of such paths is denoted with $\Omega_{p}(V)$.
\end{Definition}

The behaviour of rough paths may be described through the notion of a \emph{control}:

\begin{Definition}[Control functions, Definition 1.9 of \citet{LyonsT.J2007Dedb}]
    A \emph{control function}, or simply \emph{control}, on $[0,T]$ is a continuous non-negative function $\omega$ on $\Delta_{[0,T]}$ which is super-additive in the following sense:
    \begin{equation*}
        \omega(s, t) + \omega(t, u) \leq \omega(s, u)\quad \forall s,t,u \in [0,T]\ \text{s.t. } s\leq t\leq u
    \end{equation*}
    and $\omega(t,t) = 0\ \forall t \in[0,T]$. If, for a continuous path $X : [0,T] \to V$ and for all $(s,t) \in \Delta_{[0,T]}$, $\norm{X}{p-\mathrm{var}, [s,t]} \leq \omega(s,t)^{1/p}$ for some $p \geq 1$, then we say that the $p$-variation of $X$ is controlled by $\omega$.
\end{Definition}

A broad and useful class of rough paths -- geometric $p$-rough paths -- may be expressed as a limit of bounded variation paths in the following metric:

\begin{Definition}[The $p$-variation metric]
    Let $p \geq 1$ be a real number, and $C_{0,p}(\Delta_{[0,T]}, T^{(\lfloor{p}\rfloor)}(V))$ be the space of all continuous functions from $\Delta_{[0,T]}$ to the truncated tensor algebra $T^{(\lfloor{p}\rfloor)}(V)$ with finite $p$-variation. The $p$-variation metric between $X, Y \in C_{0,p}(\Delta_{[0,T]}, T^{(\lfloor{p}\rfloor)}(V))$ is defined as
    \begin{equation*}
        d_{p}(X, Y) = \max_{1 \leq i \leq \lfloor{p}\rfloor} \sup_{\zeta(0,T)} \left( \norm{ X^i_{t_{l-1}, t_l} - Y^i_{t_{l-1}, t_l} }{}^{\frac{p}{i}} \right)^{1/p},
    \end{equation*}
    where the supremum is taken over finite partitions $\zeta(0,T)$ of $[0,T]$.
\end{Definition}

Equipped with this metric, geometric $p$-rough paths are defined in the following way:

\begin{Definition}[Geometric $p$-rough path, Definition 3.13 of \citet{LyonsT.J2007Dedb}]
    Let $p \geq 1$ be a real number. A \emph{geometric} $p$-rough path in $V$ is a $p$-rough path that can be expressed as a limit of $1$-rough paths in the $p$-variation metric. The space of such paths is often denoted $G\Omega_{p}(V)$, and $G\Omega_{p}(V) \subset \Omega_p(V)$.
\end{Definition}

The space $G\Omega_p(V)$ of geometric $p$-rough paths is therefore the closure of $\bv{V}$ in $(\Omega_{p}(V), d_p)$ and encompasses a broad range of paths, e.g. fractional Brownian motion with Hurst parameter $> 1/4$ and continuous-time Markov processes. The following two results show that the signatures of such (geometric) $p$-rough paths are well-defined and continuous in an appropriate topology.

\begin{Theorem}[Extension Theorem, Theorem 3.7 in \citet{LyonsT.J2007Dedb}]\label{thm:ext_thm}
    Let $p \geq 1$ be a real number, $n \geq \lfloor{p}\rfloor$ an integer, and $X : \Delta_{[0,T]} \to T^{(n)}(V)$ a multiplicative functional with finite $p$-variation controlled by $\omega$. Then there exists a unique extension of $X$ to a multiplicative functional $\Delta_{[0,T]} \to T((V))$ which possesses finite $p$-variation.
\end{Theorem}

\begin{Theorem}[Continuity of the Extension Map, Theorem 3.10 in \citet{LyonsT.J2007Dedb}]
    Let $X$, $Y$ be two multiplicative functionals in $T^{(n)}(V)$ of finite $p$-variation with $n \geq \lfloor{p}\rfloor$ an integer, controlled by $\omega$. Suppose that for some $\epsilon \in (0,1)$
    \begin{equation}\label{eq:control_diff}
        \norm{X^i_{s,t} - Y^i_{s,t}}{} \leq \epsilon \frac{\omega(s,t)^{\frac{i}{p}}}{\beta \left(\frac{i}{p}\right) !}
    \end{equation}
    for $i = 1, \dots, n$ and for all $(s,t)\in\Delta{[0,T]}$. If
    \begin{equation*}
        \beta \geq 2p^2 \left\{1 + \sum_{r = 3}^{\infty} \left(\frac{2}{r - 2}\right)^{\frac{\lfloor{p}\rfloor + 1}{p}}\right\},
    \end{equation*}
    then \eqref{eq:control_diff} holds for all $i$.
\end{Theorem}

This leads us to the definition of the signature of a geometric $p$-rough path:

\begin{Definition}[The signature of a geometric $p$-rough path]
    The signature of a geometric $p$-rough path $X \in G\Omega_p(V)$ with $p$-variation controlled by some control $\omega$ is defined to be the unique extension of $X$ to a multiplicative functional in $T((V))$ under the Extension Theorem, Theorem \ref{thm:ext_thm}.
\end{Definition}

\section{Proofs}

\subsection{Proof of Proposition \ref{prop:sig_cont_abc}}\label{app:sabc_proof_a}

Here, we show that the map
    \begin{equation*}
        \mcal{D}(\bd{y}, \cdot) := \rho\left\{\sig{\bd{y}}, \cdot\right\} \circ \text{Sig} \circ \kappa\ :\ \mcal{X}^n \to \mathbb{R}_{\geq 0},\ \ \ \bd{x} \mapsto \norm{\sig{\bd{y}} - \sig{\bd{x}}}{}^2. 
    \end{equation*}
is continuous in $\bd{x}$, where $\bd{y} \in \mcal{X}^n$ is the observed dataset from the real world. We will proceed by noting that each constituent map in the above operation is a continuous map, and the result follows since compositions of continuous maps are continuous.

\begin{Lemma}\label{lem:one-var}
    Let $\mcal{X}^n$ be the space of length-$n$ basepoint-augmented sequences in $\mcal{X} = \mathbb{R}^d$ and $\bd{x}, \bd{z} \in \mcal{X}^n$. Then the one-variation
    \begin{equation}
        \onevar{\bd{x}} = \sum_{i=1}^{n-1} \norm{\bd{x}_{i+1} - \bd{x}_{i}}{\mcal{X}}
    \end{equation}
    is a norm on $\mcal{X}^n$.
\end{Lemma}
\begin{proof}
    The triangle inequality follows immediately as a result of the triangle inequality for the norm on $\mcal{X}$:
    \begin{align*}
        \onevar{\bd{x} + \bd{z}} &= \sum_{i=1}^{n-1} \norm{(\bd{x}_{i+1} + \bd{z}_{i+1}) - (\bd{x}_i + \bd{z}_i)}{\mcal{X}}\\
        &\leq \sum_{i=1}^{n-1} \norm{\bd{x}_{i+1} - \bd{x}_{i}}{\mcal{X}} + \norm{\bd{z}_{i+1} - \bd{z}_i}{\mcal{X}}\\
        &= \onevar{\bd{x}} + \onevar{\bd{z}}.
    \end{align*}
    Absolute homogeneity is also immediate:
    \begin{equation*}
        \onevar{s\bd{x}} = \sum_{i=1}^{n-1} \norm{s\bd{x}_{i+1} - s\bd{x}_i}{\mcal{X}} = |s| \sum_{i=1}^{n-1} \norm{\bd{x}_{i+1} - \bd{x}_i}{\mcal{X}} = |s| \onevar{\bd{x}}.
    \end{equation*}
    Finally, since the streams are basepoint-augmented, meaning $\bd{x}_1 = 0$ for all $\bd{x} \in \mcal{X}^n$, we have that $\onevar{\bd{x}} = 0$ iff $\bd{x} = (0, 0, \dots, 0)$:
    \begin{equation*}
        \onevar{\bd{x}} = 0\ \Longrightarrow\ \norm{\bd{x}_{i+1} - \bd{x}_i}{\mcal{X}} = 0\ \forall\ i = 1, \dots, n-1\ \Longrightarrow\ \bd{x}_i = \bd{x}_1 = 0\ \forall\ i.
    \end{equation*}
\end{proof}

We next show that lifting length-$n$ basepoint-augmented sequences in $\mcal{X}$ to sequences in $\mcal{H}$ is continuous if the canonical feature map $\phi$ associated with $\kappa$ is itself continuous:

\begin{Lemma}
Let $\mcal{X}^n$ be the space of length-$n$ basepoint-augmented sequences in $\mcal{X} = \mathbb{R}^d$, $\bd{x}, \bd{z} \in \mcal{X}^n$, and $\phi : \mcal{X} \to \mcal{H}$ be the canonical feature map associated with kernel $\kappa$ with \gls{rkhs} $\mcal{H}$. Assume $\phi$ is continuous. Then the map $\bd{x} \mapsto \kap{\bd{x}}$ -- where $\kap{\bd{x}}$ is the linear interpolation of the  points $(\phi(\bd{x}_1), \dots, \phi(\bd{x}_n))$ in $\mcal{H}$ -- is continuous in the one-variation topology.
\end{Lemma}

\begin{proof}
By Lemma \ref{lem:one-var}, the one-variation is a norm on length-$n$ basepoint-augmented sequences in $\mcal{X}$. We will proceed by showing that the one-variation is an equivalent norm to the 1-product norm, defined as
\begin{equation}
    \norm{\bd{x}}{\mcal{X}^n} := \sum_{i=1}^n \norm{\bd{x}_i}{\mcal{X}},
\end{equation}
which induces the product topology on $\mcal{X}^n$. By showing this, we will have the following implications:  from the definition of the 1-product norm,
\begin{equation}
    \norm{\bd{x} - \bd{z}}{\mcal{X}^n} < \tilde{\delta} \Longrightarrow \norm{\bd{x}_i - \bd{z}_i}{\mcal{X}} < \tilde{\delta}\ \text{also};
\end{equation}
by continuity of $\phi$, we have that $\forall\, \tilde{\epsilon} > 0$, $\exists\, \tilde{\delta} > 0$ such that
\begin{equation}
    \norm{\bd{x}_i - \bd{z}_i}{\mcal{X}} < \tilde{\delta} \Longrightarrow \norm{\phi(\bd{x}_i) - \phi(\bd{z}_i)}{\mcal{H}} < \tilde{\epsilon};
\end{equation}
and that choosing $\tilde{\epsilon} = \epsilon/2(n-1)$ for any $\epsilon > 0$ means that ensuring $\norm{\phi(\bd{x}_i) - \phi(\bd{z}_i)}{\mcal{H}} < \tilde{\epsilon}$ for all $i$ means
\begin{align}\nonumber
    \onevar{\kap{\bd{x}} - \kap{\bd{z}}} &= \sum_{i=1}^{n-1} \norm{\{\phi(\bd{x}_{i+1}) - \phi(\bd{z}_{i+1})\} - \{\phi(\bd{x}_i) - \phi(\bd{z}_i)\}}{\mcal{H}} \\\nonumber
    &\leq \sum_{i=1}^{n-1} \norm{\phi(\bd{x}_{i+1}) - \phi(\bd{z}_{i+1})}{\mcal{H}} + \norm{\phi(\bd{x}_i) - \phi(\bd{z}_i)}{\mcal{H}}\\\nonumber
    &< 2(n-1)\tilde{\epsilon}\\
    &= \epsilon.
\end{align}
We therefore have the following chain of implications: for every $\epsilon > 0$ there is a $\tilde{\delta} > 0$ such that
\begin{align}\nonumber
    \norm{\bd{x} - \bd{z}}{\mcal{X}^n} < \tilde{\delta} \Longrightarrow \norm{\bd{x}_i - \bd{z}_i}{\mcal{X}} < \tilde{\delta} \Longrightarrow \norm{\phi(\bd{x}_i) - \phi(\bd{z}_i)}{\mcal{H}} < \tilde{\epsilon}\\ 
    \quad \quad \Longrightarrow \onevar{\kap{\bd{x}} - \kap{\bd{z}}} < \epsilon.
\end{align}
It therefore suffices to show that for any $\tilde{\delta} > 0$ there is a $\delta > 0$ such that $\onevar{\bd{x} - \bd{z}} < \delta \Longrightarrow \norm{\bd{x} - \bd{z}}{\mcal{X}^n} < \tilde{\delta}$, which by this chain of implications would imply that $\forall\, \epsilon > 0$, $\exists\, \delta >0$ such that $\onevar{\bd{x} - \bd{z}} < \delta \Longrightarrow \onevar{\kap{\bd{x}} - \kap{\bd{z}}} < \epsilon$. We will do so by showing that $\onevar{\cdot}$ and $\norm{\cdot}{\mcal{X}^n}$ are equivalent norms.

We therefore seek $0 < c \leq C$ such that $c \norm{\bd{x}}{\mcal{X}^n} \leq \onevar{\bd{x}} \leq C\norm{\bd{x}}{\mcal{X}^n}$. This is trivially satisfied when $\bd{x} = (0,0,\dots,0)$, so consider $\norm{\bd{x}}{\mcal{X}^n} \neq 0$ and let $\bd{u} = \bd{x}/\norm{\bd{x}}{\mcal{X}^n}$ such that $\norm{\bd{u}}{\mcal{X}^n} = 1$. Showing that the sphere $\mcal{S} = \{ \bd{u} : \norm{\bd{u}}{\mcal{X}^n} = 1 \}$ is compact, and that $\onevar{\cdot}$ is continuous in the product topology for $\mcal{X}^n$, enables us to use the Extreme Value Theorem to find $c$ and $C$ as $\inf_{\bd{u}'\in\mcal{X}^n} \onevar{\bd{u}'}$ and $\sup_{\bd{u}'\in\mcal{X}^n} \onevar{\bd{u}'}$.

To show that $\mcal{S}$ is compact, we note that $\norm{\bd{u}}{\mcal{X}^n} = 1 \Rightarrow \norm{\bd{u}_i}{\mcal{X}} \leq 1$. The sets $\mcal{S}_i := \{ \bd{u}_i : \norm{\bd{u}_i}{\mcal{X}} \leq 1 \}$ are closed and bounded subsets of $\mcal{X} = \mathbb{R}^d$ and so are compact by the Heine-Borel Theorem. Then by Tychonoff's Theorem, the set $\prod_{i=1}^n \mcal{S}_i$ is compact under the product topology (which is induced by $\norm{\cdot}{\mcal{X}^n}$), and the sphere $\mcal{S} \subseteq \prod_{i=1}^n \mcal{S}_i$ is a closed subset of a compact set and is therefore also compact. Then, we show that $\onevar{\cdot}$ is continuous in the product topology by considering that for all $\epsilon > 0$, we have that $\forall \bd{x},\bd{z} \in \mcal{S}$
\begin{equation}
    \norm{\bd{x} - \bd{z}}{\mcal{X}^n} < \frac{\epsilon}{2} \Longrightarrow \vert{\onevar{\bd{x}} - \onevar{\bd{z}}}\vert \leq \onevar{\bd{x} - \bd{z}} \leq 2 \norm{\bd{x} - \bd{z}}{\mcal{X}^n} < \epsilon.
\end{equation}
Thus, since $\onevar{\cdot}$ is a continuous function on a compact set $\mcal{S} = \{ \bd{u} : \norm{\bd{u}}{\mcal{X}^n} = 1 \}$, then by the Extreme Value Theorem it is bounded and achieves its minimum $c=\inf_{\bd{u}'\in\mcal{X}^n} \onevar{\bd{u}'}$ and maximum $C=\sup_{\bd{u}'\in\mcal{X}^n} \onevar{\bd{u}'}$. Thus $\forall \bd{x} \in \mcal{X}^n$ with $\bd{u} := \bd{x} / \norm{\bd{x}}{\mcal{X}^n} \in \mcal{S}$,
\begin{equation}
    c \leq \onevar{\bd{u}} \leq C\ \Longrightarrow\ c\norm{\bd{x}}{\mcal{X}^n} \leq \onevar{\bd{x}} \leq C \norm{\bd{x}}{\mcal{X}^n}
\end{equation}
and so $\onevar{\cdot}$ and $\norm{\cdot}{\mcal{X}^n}$ are equivalent norms. In particular, we have that $\norm{\bd{x}}{\mcal{X}^n} \leq \onevar{\bd{x}}/c$, such that for all $\tilde{\delta} > 0$, we have that
\begin{equation}
    \onevar{\bd{x} - \bd{z}} < \delta := c\tilde{\delta}\Longrightarrow \norm{\bd{x} - \bd{z}}{\mcal{X}^n} < \tilde{\delta},
\end{equation}
and so we are done.
\end{proof}

We consider next the continuity of the signature map for piecewise linear paths of bounded variation in $\mcal{H}$. For such paths, the signature truncated at degree $1$ is a multiplicative functional with bounded variation (see \citet[][Section 3.1.2]{lyons2002system}) and, consequently, a special case of \citet[][Theorem 3.1.3]{lyons2002system} applies:
\begin{Lemma}\label{lemma:extension}
    Let $V$ be a Banach space, $x, z \in \bv{V}$ be two bounded variation paths in $V$, and $\tau$ be a constant such that
    \begin{equation*}
        \tau \geq 2 \left\{1 + \sum_{r = 3}^{\infty} \left(\frac{2}{r-2}\right)^2\right\}.
    \end{equation*}
    If $\varphi$ is a constant such that
    \begin{equation*}
        \onevar{x},\, \onevar{z} \leq \frac{\varphi}{\tau}\quad \quad \text{ and }\quad \quad \onevar{x - z} \leq \chi\, \frac{\varphi}{\tau}
    \end{equation*}
    for some $\chi > 0$, then for all $m \geq 1$
    \begin{equation}\label{eq:bound_sig_terms}
        \norm{S_m(x) - S_m(z)}{V^{\otimes m}} \leq \frac{\chi}{\tau} \cdot \frac{\varphi^m}{m!}.
    \end{equation}
\end{Lemma}

An immediate consequence of this is that the signature map is continuous in the 1-variation topology for bounded variation paths in Banach spaces:
\begin{Corollary}
    Let $\mcal{H}$ be a Hilbert space, $x, z \in \bv{\mcal{H}}$ be two bounded variation paths in $\mcal{H}$, and $\tau$ be as in Lemma \ref{lemma:extension}. 
    If $\varphi$ is a constant such that
    \begin{equation*}
        \onevar{x},\, \onevar{z} \leq \frac{\varphi}{\tau}\quad \quad \text{ and }\quad \quad \onevar{x - z} \leq \chi\, \frac{\varphi}{\tau}
    \end{equation*}
    for some $\chi > 0$, then
    \begin{equation*}
        \norm{\sig{x} - \sig{z}}{} \leq \frac{\chi}{\tau} \exp\left({\frac{\varphi^2}{2}}\right).
    \end{equation*}
\end{Corollary}
\begin{proof}
By definition of the norm on $\pH$,
    \begin{flalign*}
        && \norm{\sig{x} - \sig{z}}{} &= \sqrt{\sum_{m \geq 0} \norm{S_m(x) - S_m(z)}{\mcal{H}^{\otimes m}}^2} &&\\
        && &= \sqrt{0 + \sum_{m \geq 1} \norm{S_m(x) - S_m(z)}{\mcal{H}^{\otimes m}}^2} && (S_0(x) = 1\, \forall x \in \bv{\mcal{H}})\\
        && &\leq \sqrt{\sum_{m\geq 1} \frac{\chi^2}{\tau^2} \cdot \left(\frac{\varphi^m}{m!}\right)^2} && \text{(from  \eqref{eq:bound_sig_terms} above)}\\
        && &= \frac{\chi}{\tau} \sqrt{\sum_{m\geq 1} \frac{\left(\varphi^2\right)^m}{\left(m!\right)^2}} &&\\
        && &\leq \frac{\chi}{\tau} \sqrt{\sum_{m\geq 1} \frac{\left(\varphi^2\right)^m}{m!}} && \text{(smaller denominator)}\\
        && &\leq \frac{\chi}{\tau} \exp\left(\frac{\varphi^2}{2}\right). && \text{(convergent series)}
        \end{flalign*}
\end{proof}

We show next that the map $\rho\left(\sig{\bd{y}}, \cdot\right) : \pH \to \mathbb{R}_{\geq 0},\, s \mapsto \norm{\sig{\bd{y}} - s}{}^2$ is continuous. To do so, we make use of the following result:
\begin{Lemma}
Let $\kappa$ be a uniformly bounded kernel i.e. one for which $\sup_{x \in \mcal{X}} \sqrt{\kappa(x, x)} < \infty$, and let $\kap{\bd{x}} \in \Pl{\mcal{H}}$ be a $\mcal{H}$-valued piecewise linear path with knots at $\kap{\bd{x}_i}, i= 1, \dots, n$, and $\sig{\bd{x}}$ its signature. 
Then
\begin{equation}
    \sup_{\bd{x} \in \mcal{X}^n} \norm{\sig{\bd{x}}}{} < \infty.
\end{equation}
\end{Lemma}

\begin{proof}
For all $\bd{x} \in \mcal{X}^{n}$, we have 
\begin{flalign*}
&& \onevar{ \kap{\bd{x}} } &= \sum_{i=1}^{n-1} \norm{\kap{\bd{x}_{i+1}} - \kap{\bd{x}_{i}}}{\mcal{H}} && \text{(piecewise linear)}\\
&& &\leq \sum_{i=1}^{n-1} \norm{\kap{\bd{x}_{i+1}}}{\mcal{H}} + \norm{\kap{\bd{x}_{i}}}{\mcal{H}} && \text{(triangle inequality)}\\
&& &= \sum_{i=1}^{n-1} \sqrt{\kappa(\bd{x}_{i+1}, \bd{x}_{i+1})} + \sqrt{\kappa(\bd{x}_{i}, \bd{x}_{i})} && \text{(reproducing property)}\\
&& &\leq 2(n-1) \sup_{z \in \mcal{X}} \sqrt{\kappa(z,z)}. && \text{($\kappa$ bounded)}
\end{flalign*}
Let $v := 2(n-1) \sup_{z \in \mcal{X}} \sqrt{\kappa(z,z)}$. Then $\forall \bd{x} \in \mcal{X}^n$,
\begin{align*}
        \norm{\sig{\bd{x}}}{} &\leq \left\{\sum_{m=0}^{\infty} \frac{(\onevar{\kap{\bd{x}}}^2)^m}{\left(m!\right)^2} \right\}^{\frac{1}{2}} && \text{(Proposition \ref{thm:fact_decay})}
        \\
        &\leq \left\{\sum_{m=0}^{\infty} \frac{(v^2)^m}{m!} \right\}^{\frac{1}{2}} && \\
        &= e^{\frac{v^2}{2}}, && \text{(exponential series)}
\end{align*}
where in the first inequality we make use of the factorial decay property of signatures. We obtain the result by taking the supremum over $\mcal{X}^n$:
\begin{equation*}
    \sup_{\bd{x} \in \mcal{X}^n} \norm{\sig{\bd{x}}}{} \leq e^{\frac{v^2}{2}} < \infty.
\end{equation*}
\end{proof}

\begin{Lemma}\label{lem:ubov}
Let $\kappa$ be a uniformly bounded kernel i.e. one for which $\sup_{z \in \mcal{X}} \sqrt{\kappa(z, z)} < \infty$, and let $\kap{\bd{y}}\in \Pl{\mcal{H}}$ be the observed $\mcal{H}$-valued piecewise linear path with $\sig{\bd{y}}$ its signature. Denote the signature kernel as 
\begin{equation}
    k(\bd{x}, \bd{z}) = \langle{\sig{\bd{x}}, \sig{\bd{z}}}\rangle
\end{equation}
Then the distance function 
\begin{equation}
\rho\left(\sig{\bd{y}}, \cdot\right) : \prod_{m \geq 0}\mcal{H}^{\otimes m} \to \mathbb{R}_{\geq 0},\ \ \ s \mapsto \| s - \sig{\bd{y}}\|^2
\end{equation}
is Lipschitz continuous in $s$.
\end{Lemma}

\begin{proof}
\begin{flalign*}
&& \big\vert{ \mcal{D}(\bd{y}, \bd{x}) - \mcal{D}(\bd{y}, \bd{z}) }\big\vert &= \Big\vert{ \norm{\sig{\bd{x}} - \sig{\bd{y}}}{}^2 - \norm{\sig{\bd{z}} - \sig{\bd{y}}}{}^2 }\Big\vert &&\\
&& &= \Big\vert{ k(\bd{x}, \bd{x}) - k(\bd{z}, \bd{z}) + 2\left(k(\bd{z}, \bd{y}) - k(\bd{x}, \bd{y})\right) }\Big\vert &&\\
&& &\leq \Big\vert{ k(\bd{x}, \bd{x}) - k(\bd{z}, \bd{z}) }\Big\vert + 2\Big\vert{ k(\bd{z}, \bd{y}) - k(\bd{x}, \bd{y}) }\Big\vert && \text{(triangle inequality)}
\end{flalign*}
Considering the first of these terms and making use of the reproducing property and symmetry of $k$:
\begin{flalign*}
&& \Big\vert{ k(\bd{x}, \bd{x}) - k(\bd{z}, \bd{z}) }\Big\vert &= \Big\vert{ k(\bd{x},\bd{x}) - k(\bd{x},\bd{z}) + k(\bd{z},\bd{x}) 
 - k(\bd{z},\bd{z})\Big\vert} &\\
&& &= \Big\vert{\langle{k(\bd{x},\cdot), k(\bd{x}, \cdot) - k(\bd{z}, \cdot)\rangle} + \langle{k(\bd{z},\cdot), k(\bd{x}, \cdot) - k(\bd{z}, \cdot)\rangle}\Big\vert} && \\
&& &\leq \Big\vert{\langle{k(\bd{x},\cdot), k(\bd{x}, \cdot) - k(\bd{z}, \cdot)\rangle}}\Big\vert + \Big\vert{\langle{k(\bd{z},\cdot), k(\bd{x}, \cdot) - k(\bd{z}, \cdot)\rangle}\Big\vert} && \\
&& &\leq (\norm{\sig{\bd{x}}}{} + \norm{\sig{\bd{z}}}{}) \cdot \norm{\sig{\bd{x}} - \sig{\bd{z}}}{}, &&
\end{flalign*}
where in the penultimate and final lines we use the triangle inequality and the Cauchy-Schwarz inequality twice, respectively. Considering now the second term:
\begin{flalign*}
&& \Big\vert{ k(\bd{z}, \bd{y}) - k(\bd{x}, \bd{y}) }\Big\vert &= \Big\vert{ \langle{\sig{\bd{y}}, \sig{\bd{x}} - \sig{\bd{z}}\rangle} }\Big\vert &&\\
&& &\leq \norm{\sig{\bd{y}}}{} \norm{\sig{\bd{x}} - \sig{\bd{z}}}{}, && \text{(Cauchy-Schwartz)}
\end{flalign*}
where in the first line we use the definition and symmetry of the inner product. Putting the two terms together and using Lemma \ref{lem:ubov}, we have
\begin{align*}
    \big\vert{ \mcal{D}(\bd{y}, \bd{x}) - \mcal{D}(\bd{y}, \bd{z}) }\big\vert &\leq \left(\norm{\sig{\bd{x}}}{} + \norm{\sig{\bd{z}}}{} + 2\norm{\sig{\bd{y}}}{} \right)\norm{\sig{\bd{x}} - \sig{\bd{z}}}{}\\
    &\leq 4 e^{\frac{v^2}{2}}\norm{\sig{\bd{x}} - \sig{\bd{z}}}{}
\end{align*}
where $v$ is as in Lemma \ref{lem:ubov}. Thus $\rho\left(\sig{\bd{y}}, \cdot\right)$ is Lipschitz continuous.
\end{proof}

We finally arrive at the conclusion:

\begin{Proposition}
    The map
    \begin{equation}
        \mcal{D}(\bd{y}, \cdot) := \rho\left\{\sig{\bd{y}}, \cdot\right\} \circ \text{Sig} \circ \kappa : \mcal{X}^n \to \mathbb{R}_{\geq 0},
    \end{equation}
    consisting of lifting the sequence $\bd{x} \in \mcal{X}^n$ to a piecewise linear path in $\mcal{H}$, before computing the squared distance between its signature and $\sig{\bd{y}}$, is uniformly continuous.
\end{Proposition}
\begin{proof}
Compositions of continuous maps are continuous, and each of the constituent maps are continuous from the Lemmas and Corollaries presented above.
\end{proof}

\subsection{Proof of Proposition \ref{prop:sig_inj_abc}}\label{app:sabc_proof_b}

\begin{proof}
    Obtaining a signature from a length-$n$ data stream $\bd{x}$ entails: (1) lifting the points $\bd{x}_i$ in $\bd{x}$ to the \gls{rkhs} $\mcal{H}$ associated with $\kappa$ as $\kap{\bd{x}_i}$; (2) applying a linear interpolation to obtain a piecewise linear $\mcal{H}$-valued path $\kap{\bd{x}}$; and (3) finally taking the signature of $\kap{\bd{x}}$. To show injectivity of this composite map, it suffices to show injectivity of each of these three steps since the composition of injective maps is injective.
    
    (1) is trivially injective, due to the assumed injectivity of $\kappa$. (2) is by definition injective for a length-$n$ sequence in $\mcal{H}$. To show injectivity of (3), we note that time-augmentation of the sequences, along with injectivity of $\kappa$, ensure that the lifted paths are injective, such that no tree-like equivalence is observed between the interpolated paths in $\mcal{H}$. Time-augmentation further makes the signature sensitive to parameterisation, removing its parameterisation invariance property. Uniform boundedness of $\kappa$ ensures that $\kap{\bd{x}}$ is of bounded variation, such that $\kap{\bd{x}} \in \Pl{\mcal{H}}$. To see this, note that for a piecewise linear path $\kap{\bd{x}}$,
    \begin{equation*}
        \onevar{\kap{\bd{x}}} = \sum_{i = 1}^{n-1} \norm{\kap{\bd{x}_{i+1}} - \kap{\bd{x}_{i}}}{\mcal{H}} \leq 2(n-1)\sup_{\bd{z} \in\mcal{X}} \sqrt{\kappa(\bd{z}, \bd{z})} < \infty,
    \end{equation*}
    where we have used the reproducing property of $\kappa$ and the triangle inequality. Finally, since basepoint augmentation makes the signature sensitive to paths that differ only by translations, the desired result follows from Theorem \ref{thm:sig_inj_tle}.
\end{proof}

\subsection{Proof of Proposition \ref{prop:mesh_prop}}\label{app:mesh_proof}

Throughout this section, we will denote with $\zeta(0,T)$ a partition of the interval $[0,T]$, $\Delta_{[0,T]} := \lbrace{(s,t) \in [0,T]^2 : 0 \leq s \leq t \leq T\rbrace}$, $\mesh$ the largest interval in $\zeta(0,T)$ i.e.
    \begin{equation*}
        \mesh := \max_{(s,t) \in \zeta(0,T)} | t - s |
    \end{equation*}
with $0 \leq s \leq t \leq T$, 
\begin{equation*}
    \mcal{D}(\bd{x}, \bd{y}) = \norm{\sig{\bd{x}} - \sig{\bd{y}}}{}^2,\quad \quad \bd{x}, \bd{y} \in \Pl{\mcal{X}}
\end{equation*}
and
\begin{equation*}
    \mcal{D}(h, g) = \norm{\sig{h} - \sig{g}}{}^2,\quad \quad h, g \in \bv{\mcal{H}}.
\end{equation*}

\begin{Lemma}\label{lem:dist_converge}
    Let $\kappa$ be a uniformly bounded, injective kernel on $\mcal{X}$, and $\kap{\bd{x}}, \kap{\bd{y}} \in \Pl{\mcal{H}}$ be the simulated and observed datasets, respectively, which are discretisations of underlying paths $h,g \in \bv{\mcal{H}}$. Then,
    \begin{equation}
        \norm{\sig{\bd{x}} - \sig{\bd{y}}}{}^2 \longrightarrow
        \norm{\sig{h} - \sig{g}}{}^2
    \end{equation}
    as $\mesh \to 0$.
\end{Lemma}

\begin{proof}
    Let $\rho = (s_i)_{i=1}^N, 0 = s_1 < \dots < s_N = T$ and $\varrho = (t_j)_{j=1}^{M}, 0 = t_1 < \dots < t_M = T$ be partitions of the interval $[0,T]$ such that $\kap{\bd{x}}_{s_i} = h_{s_i}, i = 1, \dots, N$ and $\kap{\bd{y}}_{t_j} = g_{t_j}, j = 1, \dots, M$, with the paths $\kap{\bd{x}}, \kap{\bd{y}} \in \Pl{\mcal{H}}$ linear in between these points. Then,  by \citet[][Corollary 4.7]{Kiraly2019},
    \begin{equation*}
        \vert{ k(\bd{x}, \bd{y}) - k(h, g)} \vert \leq 2e^{\onevar{h} + \onevar{g}} - e^{\onevar{h} + \onevar{\kap{\bd{y}}}} - e^{\onevar{\kap{\bd{x}}} + \onevar{g}},
    \end{equation*}
    where convergence is uniform. 
    Therefore, 
    \begin{flalign*}
        && \big\lvert{\mcal{D}(\bd{x}, \bd{y}) - \mcal{D}(h, g)}\big\rvert &= \big\lvert{ k(\bd{x}, \bd{x}) - k(h, h) + k(\bd{y}, \bd{y}) - k(g, g) + 2(k(h, g) - k(\bd{x}, \bd{y})) }\big\rvert &&\\
        && &\leq \big\lvert{ k(\bd{x}, \bd{x}) - k(h, h)}\big\rvert + \big\lvert{k(\bd{y}, \bd{y}) - k(g, g)}\big\rvert + 2\big\lvert{k(h, g) - k(\bd{x}, \bd{y}) }\big\rvert && 
        \\
        && &\longrightarrow 0\ \,\text{  as  }\ \mesh \longrightarrow 0, && 
    \end{flalign*}
    where the triangle inequality is used in the second line.
\end{proof}

We may now state a proof of Proposition \ref{prop:mesh_prop}:

\begin{proof}[Proof of Proposition \ref{prop:mesh_prop}]
    By Lemma \ref{lem:dist_converge}, $\mcal{D}(\bd{x}, \bd{y}) \to \mcal{D}(h, g)$ where $h, g \in \bv{\mcal{H}}$ are the bounded variation paths of which $\kap{\bd{x}}$ and $\kap{\bd{y}}$ are discretisations. Further, by our choice of $\varepsilon$, $\mathbb{P}\{\mcal{D}(h,g) = \varepsilon\} = 0$ and $\mathbb{P}\{\mcal{D}(h,g) < \varepsilon\} > 0$, where $\mathbb{P}$ denotes a probability measure. Then, we follow \citet{miller2018robust} and apply Lemma 5.1 contained therein using the same notation: we obtain the result by taking the ordered sequence $(U_n : n \geq 1)$ to be the $\mcal{D}(\bd{x}, \bd{y})$ as $\mesh$ decreases and $\onevar{\kap{\bd{x}}}, \onevar{\kap{\bd{y}}} \to \onevar{h}, \onevar{g}$; $U = \mcal{D}(h, g)$; $V = \varepsilon$; and $W = h(\bth)$ for any continuous, bounded $h : \bTh \to \mathbb{R}$.
\end{proof}

\section{Further experimental details}

\subsection{Signature Regression ABC}\label{app:skabc}

For \gls{skrr}, we proceed as follows: 
\begin{enumerate}
    \item[(a)] fit a kernel ridge regression model using training data $\lbrace{\bx^{(i)}, \bth^{(i)}\rbrace}_{i=1}^{R} \sim p\left(\bx, \bth\right)$. This amounts to solving the following optimisation problem for each of the $p$ components $j=1,\dots,p$ of the $\lbrace{\bth^{(i)}\rbrace}_{i=1}^{R}$:
    \begin{equation}
        \min_{\hat{\bth}_{j}\in \mathcal{H}_k} \sum_{i=1}^{R} \left\{\bth_j^{(i)} - \hat{\bth}_{j}\left(\bx^{(i)} \right)\right\}^2 + \alpha \| \hat{\bth}_{j} \|^{2}_{\mathcal{H}_k},
    \end{equation}
    where $k$ is the signature kernel, $\mathcal{H}_k$ is the \gls{rkhs} associated with $k$, $\hat{\bth}_{j}$ is -- by the Representer Theorem -- a function of the form
    \begin{equation}
        \hat{\bth}_{j}(\bx) = \sum_{i=1}^{R} \boldsymbol{\omega}^{(j)}_i k(\bx, \bx^{(i)})
    \end{equation}
    with
    \begin{gather*}
        \boldsymbol{\omega}^{(j)} = \left(G + \alpha I_R\right)^{-1}\bpsi^{(j)},\quad \quad \quad G_{mn} = k(\bx^{(m)}, \bx^{(n)}),\\[1ex] %
        \bpsi^{(j)} = 
            \left[
            \begin{array}{c}
                \bth_{j}^{(1)}\\ \bth_{j}^{(2)}\\ \vdots\\ \bth_{j}^{(R)}
            \end{array}
            \right],
            \quad \quad I_R = \text{diag}(1, 1, \dots, 1) \in \mathbb{R}^{R\times R},
    \end{gather*}
    and $\alpha \geq 0$ is a regularisation parameter;
    \item[(b)] summarise the observation $\by$ and all future simulations $\bx\sim p(\bx \mid \bth)$ using this trained kernel ridge regression model, i.e. use
    \begin{equation}
        \bs(\bx) = \left[
        \begin{array}{c}
            \hat{\bth}_{1}\left(\bx\right)\\ \hat{\bth}_{2}\left(\bx\right)\\ 
            \vdots \\ 
            \hat{\bth}_{p}\left(\bx\right)
        \end{array}\right];
    \end{equation}
    \item[(c)] use the squared difference between the summaries of $\by$ and $\bx$ as the measure of discrepancy between simulation and observation,
    \begin{equation}
        \rho\left\{ \bs(\by), \bs(\bx)\right\} = \| \bs(\by) - \bs(\bx) \|^2_2.
    \end{equation}
\end{enumerate}

\subsection{Further implementation details}\label{app:imp_det}

For all signature kernel computations, we use the \texttt{sigkernel} package \citep{salvi2020computing} and we normalise the time series by %
dividing by the range of the simulation output when this is known or, when this is unknown, with the expected range of the training set of size $R=300$ for \gls{skrr} or $R=300$ samples from the prior predictive distribution for \gls{sabc}. %

Unless stated otherwise, we remove the translation invariance and reparameterisation-invariance properties of the signature -- discussed in Section \ref{sec:invariances} -- by applying basepoint and time-augmentations to all time series in every experiment.

Unless stated otherwise, we take $\kappa$ to be a Gaussian RBF kernel with scale hyperparameter $\sigma$. To tune $\sigma$ and the regularisation hyperparameter for \gls{skrr}, we perform a grid search with 5-fold cross-validation on the training set. For \gls{sabc}, we use the median of all pairwise Euclidean distances between points in the observation $\by$ for $\sigma$, although we note that other approaches could be taken, such as using the same method as for \gls{skrr}. 

Both \gls{saabc} and \gls{skrr} %
require training data; for both we use $R = 300$ training examples $\lbrace{\bx^{(j)}, \bth^{(j)}}\rbrace_{j=1}^{R} \sim p(\bx \mid \bth)\pi(\bth)$. When $\pi(\cdot)$ has bounded support, we normalise the parameters $\lbrace{\bth^{(i)}\rbrace}_{i=1}^{R}$ in the training set with the range of the prior in each dimension. We also tune the bandwidth parameter for the Gaussian RBF kernel employed in the \gls{mmd} loss for \gls{k2abc} using the median of the pairwise absolute differences between observations in $\by$, as recommended by \citet{Park2016}. 

In all experiments, \gls{wass} indicates the 1-Wasserstein distance with curve matching, which as described in Section \ref{sec:Back} is a method for using the Wasserstein distance for time series recommended in \citet{Bernton2019}. To determine the $\lambda$ coefficient, we follow the guidance of \citet{thorpe2017transportation} and choose
\begin{equation}
    \lambda \simeq \frac{V}{T},
\end{equation}
where $V$ is the expected vertical range and $T$ is the length of the time interval over which observations are made, in order to balance the effects of vertical and horizontal transport. Where the value of $V$ is not apparent \emph{a priori}, we estimate it using $R=300$ samples from the prior predictive distribution. Distances are computed using the Python Optimal Transport package \citep{flamary2021pot}.

\subsection{Reference Posteriors using MCMC}
\label{app:mcmc}
\glsresetall

\paragraph{Metropolis-Hastings} For the \gls{gbm} and Brock \& Hommes models, we obtain samples from the ground truth posterior using \gls{mh}. We follow the guidelines of \citet{schmon2021optimal} and use a multivariate normal proposal, for which we estimate the covariance matrix using a pilot run. We subsequently tune the \gls{mh} algorithm according to \citet[][Table 1]{schmon2021optimal} and run the \gls{mh} for $10^5$ steps, keeping a thinned subset of $10^3$ samples as our baseline.

\paragraph{Particle MCMC} To obtain samples from the ground truth posterior of the Ricker model we employ \gls{pmcmc} using a simple bootstrap particle filter. 
We follow the guidelines of \citet{SchmonDeligiannidisDoucet2018a}, first estimating the posterior covariance  in a shorter prior run and then tuning the random walk proposal as well as the particle filter. 
\Gls{pmcmc} commonly exhibits worse convergence behaviour than standard \gls{mh} and hence we run the algorithm for $2\times 10^5$ iterations eventually retaining a thinned subset of $10^3$ samples as our baseline.

\subsection{Example code for Signature ABC}\label{app:sabc_code}

The distance function \eqref{eq:sig_distance} can be computed easily with the \texttt{sigkernel} package \citep{salvi2020computing}. We offer the following as an example:

\begin{lstlisting}[language=Python, caption=Example python code for computing the distance between signatures.]
import model # The simulator
import sigkernel # For computing the signature kernel

# Generate observation
y = model.simulate()

# Specify static kernel, which is sequentialised in the signature kernel
sigma_y = median_heuristic(y)
static_kernel = sigkernel.RBFKernel(sigma=sigma_y)

# Choose the dyadic order for the finite element PDE solver (integer,
# default is 0, higher values give more accurate PDE solutions but 
# are more expensive. dyadic_order = 1 is taken in all experiments above)
dyadic_order = 1

# Sequentialise the above static kernel to create a signature kernel
signature_kernel = sigkernel.SigKernel(static_kernel, dyadic_order)
k = signature_kernel.compute_kernel

# Compute distance between simulation and observation
x = model.simulate()
distance = k(x, x) + k(y, y) - 2*k(x, y)
\end{lstlisting}

\end{document}